\documentclass[aps,prb,groupedaddress,twocolumn]{revtex4}
\usepackage{epsfig}
\usepackage[dvipsnames,usenames]{color}
\usepackage{color}
\usepackage{amsmath}
\usepackage{comment}
\usepackage{array}
\usepackage{multirow}
\usepackage{tabularx}
\usepackage{float}

\emergencystretch=\maxdimen
\begin{document}
\title{Relevance of Cu-3d multiplet structure in models of high Tc cuprates}
\author{Mi Jiang, Mirko Moeller, Mona Berciu, and George A. Sawatzky}
\affiliation{Department of Physics and Astronomy, University of
  British Columbia, Vancouver B.C. V6T 1Z1, Canada}
\affiliation{Stewart Blusson Quantum Matter Institute, University of
British Columbia, Vancouver B.C. V6T 1Z4, Canada}

\begin{abstract}
We revisit the problem of the spectra of two holes in a
CuO$_{2}$ layer, modeled as a Cu-d$^{8}$ impurity with full multiplet
structure coupled to a full O-2p band as an approximation to the local electronic structure of a hole doped cuprate. Unlike previous studies
that treated the O band as a featureless bath, we describe it with a
realistic tight binding model. While our results are in qualitative
agreement with previous work, we find considerable quantitative
changes when using the proper O-2p band structure. We also find (i)
that only the ligand O-2p orbitals play an essential role, within this
impurity model; (ii) that the three-orbital Emery model provides an
accurate description for the subspace with $^{1}\!A_1$ symmetry, which
includes the ground-state in the relevant region of the phase diagram;
(iii) that this ground-state has only $\sim 50\%$ overlap with a
Zhang-Rice singlet; (iv) that there are other low-energy states, in
subspaces with different symmetries, that are absent from the
three-orbital Emery model and its one-band descendants. These states play an important role in describing the elementary excitations of doped cuprates. 
\end{abstract}

\maketitle

\section{Introduction}
A central issue still under debate in the study of high-$T_{c}$
cuprate superconductors is the proper minimal model that correctly
captures the low-energy properties, specifically  the precise nature of
the states closest to the Fermi level. Historically, Anderson proposed
that the essential physics can be understood based on the single-band
Hubbard model, where the band is identified as the antibonding band of
Cu-$3d_{x^2-y^{2}}$ and O-$2p$ orbitals.\cite{Anderson87} The even
simpler t-J model additionally discards all doubly occupied states and
describes a square lattice where charge carriers move in a spin
background. This has been extensively studied away from half-filling
and is believed to provide a good description of the Hubbard model in
the strong coupling limit $J/t = 4t/U \ll 1$. However, their common
intrinsic assumption is that the cuprate parents compounds, which are
known to be charge transfer insulators\cite{GSA1985}, can instead be modeled as
effective Mott-Hubbard insulators.

The need to understand the importance of explicitly including the O
ions hosting the doped holes, motivated the study of the three-band
Emery model\cite{Emery}, which includes the Cu $d_{x^2-y^{2}}$ and the
two ligand 0-2p$_{\sigma}$ orbitals in the non-magnetic unit cell. The key idea underlying the expected
equivalence of the one- and three-orbital scenarios was proposed by
Zhang and Rice, who argued that the doped holes occupy a certain
linear combination of O orbitals that is locked into a Zhang-Rice
singlet (ZRS) with the hole (spin) residing on the central Cu site.
Projecting onto these ZRS then allows one to map the three-band Emery
model onto a single-band t-J model,\cite{ZhangRice} although a more
careful treatment reveals the existence of additional terms 
ignored by the t-J Hamiltonian.\cite{Aligia1,Aligia2}

Although various analytical approximations and extensive numerical
studies of these model Hamiltonians have revealed many insights in the
past decades, the validity of the ZRS
concept\cite{nonZhangRice2017,T-CuO} and more generally the
equivalence -- or lack thereof -- between the low-energy properties of
one- and three-orbital models are still under debate. On one hand, the
existence and stability of states with ZRS-like character have been
confirmed in previous photoemission
experiments.\cite{CuO,CuO_ZRS,BSCCO_ZRS} On the other hand, recent
calculations contrasting the dynamics of a single doped hole in the
one-band {\em vs.} the three-band model revealed qualitative
differences,\cite{Lau,Hadi1, Hadi2} such as the essential
{\em vs.} the minor role played by the background spin-fluctuations,
respectively. Moreover, a recent high-energy optical conductivity
study questioned the ZRS argument by revealing a strong mixture of
singlet and triplet configurations in the lightly hole-doped Zn-LSCO
single crystal.\cite{nonZhangRice2017} Furthermore, this system exhibits
strong ferromagnetic correlations between Cu spins near the doped
holes, as predicted by the three-band model.\cite{Lau}

During the same period when the ZRS was proposed, Eskes {\it et al.}
carried out a more general study that included the multiplet structure
of the Cu, {\em i.e.} all singlet and triplet irreducible
representations in the $D_{4h}$ point group spanned by two d holes
($d^8$-type configurations) and their corresponding Coulomb and
exchange interactions\cite{Zaanen1987,Eskes88,Eskes90}, besides explicitly
considering the O band. This was achieved at the cost of simplifying
the model to consist of a single Cu impurity hybridizing with a broad
O band described in terms of a featureless, semiellipical density of
states.

This work confirmed that the first ionization state starting from a Cu-$d^9$ state and a full O-$2p$ band, which ends with the two hole eigenstates involving $d^8$ multiplets and various continuum states, is indeed in the $^{1}A_1$ symmetry channel consistent with the symmetry of the ZRS,
but also found that the energy difference between the lowest
ionization states for various symmetry channels is rather small.
Moreover, these differences are strongly dependent on the electronic
structure, which in turn is likely to depend quite strongly on doping
levels. These results cast doubt on whether it suffices to include
only the $d_{x^2-y^2}$ orbital instead of the full 3d multiplet
structure of the Cu-$d^8$, when modeling these materials.

Most members of our community believe that the Cu-$d_{x^2-y^{2}}$
orbital is the only d-orbital needed to account for the essential
physics of cuprates, explaining why there are so few studies on the
effects of the multiplet structure, compared to the very extensive
investigations of the one- and three-band models involving only Cu
$d_{x^2-y^{2}}$ orbital and/or its ZRS daughter. However, there are
both theoretical and experimental results pointing out the importance of non-planar
orbitals like Cu-$3d_{3z^2-r^2}$ and/or
O-$2p_z$\cite{Feiner1992,Feiner1992a,Feiner1992b,Feiner1992c,Feiner1992d,Feiner1992e,Feiner1992f,
Feiner1992g,Tjeng2003,HideoAoki2012,arpes2018,Jin2018}. In particular, the
importance of Cu-$3d_{3z^2-r^2}$ is revealed by the recent
discovery\cite{Jin2018} of the cuprate superconductor
Ba$_2$CuO$_{4-\delta}$ with critical temperature $T_c\sim 70$ K, where based on the compresed c-axis bond length, it is claimed that some doped holes are likely in the $d_{3z^2-r^2}$ orbital.
 Early Auger spectroscopic experiments~\cite{Auger1980} clearly demonstrated strong multiplet effects ranging over a large energy scale in Cu compounds such as CuO and Cu$_2$O. In fact, in Cu$_2$O the lowest energy Cu $d^8$ state is a triplet state consistent with the Hund's rule expectations. As pointed out by Eskes\cite{Eskes88,Eskes90}, the crossing of the singlet and triplet states in the cuprate parent compounds is a result of the strong O character in these states due to the strong Cu-O hybridization and the fact that the Cu $d^8$ states are pushed out of the top of the O 2p band resulting in the lowest energy singlet bound states\cite{CuO,CuO_ZRS,Auger1980}.

Also important evidence for the significant role of the multiplets comes from X-ray
absorption (XAS) experiments that have shown, upon increased doping, a
strong change from purely $x,y$ polarized absorption to one including
a large contribution of $z$ polarized intensity for the O and Cu
core-to-valence transition~\cite{Feiner1992}. This implies that there are doped holes
whose wavefunctions have a considerable component in the
Cu-$d_{3z^2-r^2}$ or O-$p_z$ orbitals. These results point to the
breakdown of the single-band or even three-band (Cu-$d_{x^2-y^{2}}$
based) approaches to the description of the phase diagram of cuprate
superconductors. 

This motivates us to revisit the importance of the full Cu-$3d$
multiplet structure and explore its effects on the low-energy
properties of cuprate models. In order to obtain numerically exact
results, we follow Eskes {\it et al.} and study a single Cu impurity
with all its 3d orbitals included. In contrast to this earlier work,
however, we properly embed this Cu impurity in a square lattice of O
2p orbitals, with a realistic band-structure. This allows us to
contrast models containing only the O-2p ligand orbitals {\em vs.}
those also including the other in-plane orbital, and also the $p_z$
orbital. The results reveal the importance of the realistic modeling of
the O bath, and which O-2p orbitals play an essential role.
It is important to note that the linear combination of O 2p orbitals that hybridize with the various Cu 3d states live in different energy regions of the O 2p band structure and this strongly influences the importance of this hybridization. For example, the $d_{xz}$ orbital hybridizes with the O-$p_z$ and $p_x$ $\pi$-bonding orbitals while the $d_{x^2-y^{2}}$ orbital hybridizes with the O-2p $\sigma$-bonding orbitals. Besides, the linear combination of the O-2p orbitals that hybridize with $d_{x^2-y^{2}}$ orbital is different in their relative phases than with the $d_{3z^2-r^2}$ orbital. We will see below how this strongly influences the appearance and relative energies of bound states pushed out of the O band for the various symmetries.
Furthermore, by calculating the Cu-3d electron removal spectra in
various symmetry channels of the $D_{4h}$ point group, we are able to
identify the character (symmetry, spin, and orbital composition) of
the first ionization state, and to gauge its similarity to a ZRS.
Finally, our results reveal strong similarities between the model
including all multiplets and the conventional three-orbital Emery
model if we restrict ourselves to the lowest energy electron removal states, although open issues still remain.
However, if one wants to describe the spectroscopies like ARPES going up to one or more eV below the Fermi energy as, for example, in descriptions of the so called ``waterfall'' feature~\cite{Lanzara}, it is essential to include all of the multiplets since they all have appreciable spectral weights extending to energies well above 1 eV. 

This paper is organized as follows. In Sec.~\ref{model}, we define our
model and the variational method employed to study its single-doped
hole eigenstates. Sec.~\ref{results} discusses the resulting spectra
for various cases considered. The summary and future issues to be
addressed are presented in Sec.~\ref{Conclusion}.

\section{Models and Methods}\label{model}

\subsection{Multi-orbital models with a single Cu impurity}

\begin{figure}[t]
\psfig{figure=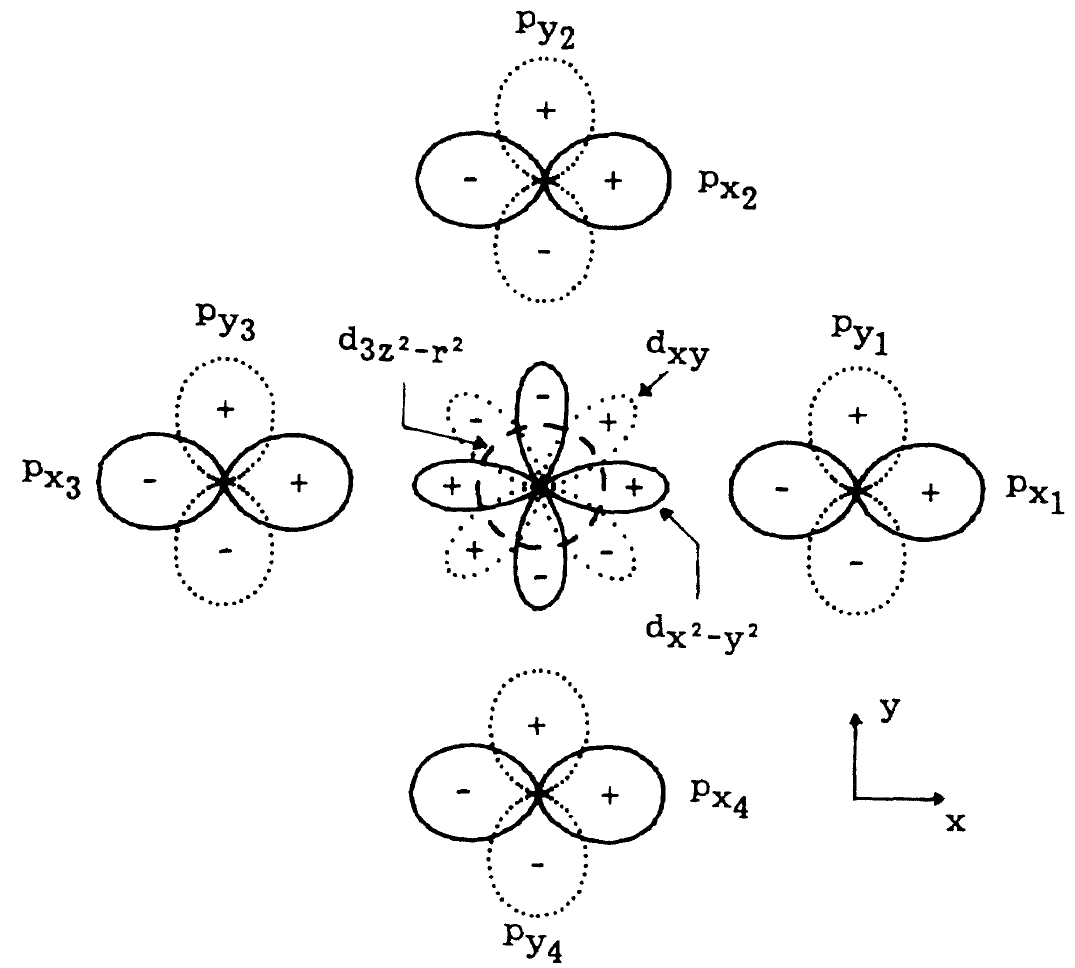,height=6.5cm,width=7.0cm,angle=0,clip} \\
\caption{Schematic view of the orbitals involved in our model calculations, adapted from Eskes's previous related work.\cite{Eskes90} The Cu $d_{xz}, d_{yz}$ and the O $p_{z}$ orbitals are not shown. {Note that only the four O that are adjacent to the Cu impurity are depicted, however we consider the full O square lattice.} }
\label{orbs}
\end{figure}

We simplify the description of a CuO$_2$ plane by replacing the Cu
lattice with a single Cu impurity properly embedded in a square lattice of O
orbitals; the resulting problem can be solved exactly, unlike the
corresponding one for the full CuO$_2$ lattice. The central part of
the system, consisting of the Cu impurity and its 4 nearest neighbor
(NN) O ions, is depicted in Fig.~\ref{orbs}. The Hamiltonian describing this system is
\begin{align} \label{H}
H &= E_{s} + K_{pd} + K_{pp} + V_{dd} + V_{pp} \nonumber \\
E_s &= \sum_{m\sigma} \epsilon_d(m) d^\dagger_{m\sigma}d^{\phantom\dagger}_{m\sigma} 
      + \sum_{jn\sigma} \epsilon_p p^\dagger_{jn\sigma}p^{\phantom\dagger}_{jn\sigma} \nonumber \\ 
K_{pd} &= \sum_{\langle .j\rangle mn \sigma} 
(T^{pd}_{mn} d^\dagger_{m \sigma}p^{\phantom\dagger}_{jn \sigma}+h.c.) \nonumber \\ 
K_{pp} &= \sum_{\langle jj'\rangle nn' \sigma} 
(T^{pp}_{nn'} p^\dagger_{jn \sigma}p^{\phantom\dagger}_{j'n' \sigma}+h.c.) \nonumber \\ 
V_{dd} &= \sum_{\bar{m}_1\bar{m}_2\bar{m}_3\bar{m}_4} U(\bar{m}_1\bar{m}_2\bar{m}_3\bar{m}_4) d^\dagger_{\bar{m}_1}d^{\phantom\dagger}_{\bar{m}_2}d^\dagger_{\bar{m}_3}d^{\phantom\dagger}_{\bar{m}_4}
\end{align}
Here, the simplified notation $\bar{m}_x \equiv m_x \sigma_x$ in $V_{dd}$ with $x=1,2,3,4$ denote the spin-orbital. The  $E_{s}$ represents the onsite
energies, where $d^\dagger_{m\sigma}(d^{\phantom\dagger}_{m\sigma})$ creates (destroys) a
hole in the Cu-3d orbital $m$ with on-site energy $\epsilon_d(m)$ and spin
$\sigma$, while $p^\dagger_{jn\sigma}(p^{\phantom\dagger}_{jn\sigma})$ creates (destroys) a hole
at the O lattice site $j$, in its 2p orbital $n$ with energy $\epsilon_p$ and
spin $\sigma$. The Cu-3d orbitals indexed by $m$ are $b_1(d_{x^2-y^{2}}),
a_1(d_{3z^2-r^{2}}), b_2(d_{xy}), e_x(d_{xz}), e_y(d_{yz})$, and the
O-2p orbitals indexed by $n$ are $p_x, p_y, p_z$ or a subset of them,
as indicated below. All other core levels and the Cu 4s and 4p
orbitals are neglected because of their high-energy, which
allows for their influence via hybridization to be accounted for by
the renormalization of the effective parameters. Finally, the onsite
d-hole energies $\epsilon_d(m)=0$ are assumed to be independent of $m$, thus
omitting the point-charge crystal splitting. This is expected to be a
good approximation because it is the hybridization with the O
orbitals, included in our model, that accounts for most of the
difference between the effective on-site energies of the 3d levels.
As a result, the charge-transfer energy $\Delta=\epsilon_p$.

$K_{pd}$ and $K_{pp}$ describe the Cu-O and O-O hoppings,
respectively. The labels $j,j'$ run over the positions of the O atoms,
$\langle .j\rangle$ is a sum over the four O adjacent to the $j^{th}$ Cu site, and only NN
$pp$ hopping is included. Following Slater and
Koster,\cite{SlaterKoster} the Cu-O and O-O hopping integrals
$T^{pd}_{mn}$ and $T^{pp}_{nn'}$ are listed in Table~\ref{table1}. Throughout the paper, energies are measured in eV.

In the following we focus on four possible models: (i) N3, where $m=b_1$
and $n\in \{p_{x_1},p_{y_2} \}$, {\em i.e.} the usual three-band Emery
model where only the ligand orbital is kept for each O; (ii) N7, where
$m\in\{a_1, b_1, b_2, e_x, e_y\}$ and $n\in \{p_{x_1},p_{y_2} \}$, {\em i.e.} multiplet-like physics is added to the Emery model; and
(iii) N9, where $m\in\{a_1, b_1, b_2, e_x, e_y\}$ and $n \in
\{p_{x_1},p_{y_1},p_{x_2},p_{y_2} \}$, {\em i.e.} for each O we keep
both in-plane 2p orbitals; and (iv) N11, where $m\in\{a_1, b_1, b_2, e_x, e_y\}$ and $n \in
\{p_{x_1},p_{y_1},p_{z_1},p_{x_2},p_{y_2},p_{z_2} \}$, {\em i.e.} for each O we keep
all three O-2p orbitals. 

For the N9 and N11 models we use
$T^{pd}_{b_2}=T^{pd}_{b_1}/2$, so that $t_{pd\pi} = \sqrt{3} t_{pd\sigma}/4$. We
emphasize that all the Cu-O hybridization parameters $t_{pd}, t_{pp}, t_{pd\sigma}, t_{pd\pi}, t_{pp\sigma}, t_{pp\pi}$ are taken to be positive, and the signs due to the
orbitals' overlap (see Fig.~\ref{orbs}) are explicitly indicated in Table~\ref{table1}.

\begin{table}[b]
\footnotesize
\caption{The Cu-O and O-O hopping integrals $T^{pd}_{mn}$ and $T^{pp}_{nn'}$ with $m \in \{b_1(d_{x^2-y^{2}}), a_1(d_{3z^2-r^{2}}), b_2(d_{xy}), e_x(d_{xz}), e_y(d_{yz}) \}$ for various models. The hoppings involving $p_{x_3},p_{y_3},p_{x_4},p_{y_4}$ follow the sign convention illustrated in Fig.~\ref{orbs}.}
\centering
\begin{tabular}{c|c c|c c|c c c c} 
 \hline\hline
 \multirow{2}{*}{$m$} & \multicolumn{2}{c|}{N3} & \multicolumn{2}{c|}{N7} & \multicolumn{4}{c}{N9} \\
 \cline{2-9}
    & $T^{pd}_{mx_1}$ & $T^{pd}_{my_2}$ & $T^{pd}_{mx_1}$ & $T^{pd}_{my_2}$ & $T^{pd}_{mx_1}$ & $T^{pd}_{my_1}$ & $T^{pd}_{mx_2}$ & $T^{pd}_{my_2}$ \\ [0.5ex] 
 \hline
 $b_1$ & -$t_{pd}$ &  $t_{pd}$ & -$t_{pd}$ &  $t_{pd}$ & -$\sqrt{3} t_{pd\sigma}/2$ & $0$ & $0$ & $\sqrt{3} t_{pd\sigma}/2$ \\
 $a_1$ &    &    & $t_{pd}/\sqrt{3}$ & $t_{pd}/\sqrt{3}$ & -$t_{pd\sigma}/2$ & $0$ & $0$ & -$t_{pd\sigma}/2$ \\ 
 $b_2$ &    &    &    &    & $0$ & $t_{pd\pi}$ & $t_{pd\pi}$ & $0$ \\
 \hline\hline
\end{tabular} \\
\begin{tabular}{c|c c c c c c} 
 \multirow{2}{*}{$m$} & \multicolumn{6}{c}{N11} \\
 \cline{2-7}
    & $T^{pd}_{mx_1}$ & $T^{pd}_{my_1}$ & $T^{pd}_{mz_1}$ & $T^{pd}_{mx_2}$ & $T^{pd}_{my_2}$ & $T^{pd}_{mz_2}$ \\ [0.5ex] 
 \hline
 $b_1$ & -$\sqrt{3} t_{pd\sigma}/2$ & $0$ & $0$ & $0$ & $\sqrt{3} t_{pd\sigma}/2$ & $0$ \\
 $a_1$ & -$t_{pd\sigma}/2$ & $0$ & $0$ & $0$ & -$t_{pd\sigma}/2$ & $0$ \\ 
 $b_2$ &  $0$ & $t_{pd\pi}$ & $0$ & $t_{pd\pi}$ & $0$ & $0$ \\
 $e_x$ &  $0$ & $0$ & $t_{pd\pi}$ & $0$ & $0$ & $0$ \\
 $e_y$ &  $0$ & $0$ & $0$ & $0$ & $0$ & $t_{pd\pi}$ \\
 \hline\hline
\end{tabular} \\
\centering
\begin{tabular}{c|c c c c} 
\centering
 N3/N7 & \multicolumn{4}{c}{N9/N11} \\
 \hline
 $T^{pp}_{x_1 y_2}$ & 2$T^{pp}_{x_1 x_2}$ &  2$T^{pp}_{x_1 y_2}$ & 2$T^{pp}_{x_2 y_1}$ & 2$T^{pp}_{y_1 y_2}$ \\ [0.5ex] 
 \hline
 $t_{pp}$ & $t_{pp\pi}-t_{pp\sigma}$ & $t_{pp\pi}+t_{pp\sigma}$ & $t_{pp\pi}+t_{pp\sigma}$ & $t_{pp\pi}-t_{pp\sigma}$ \\ 
 \hline\hline
\end{tabular}
\label{table1}
\end{table}

In this impurity model, the single electron removal eigenstates of the undoped Cu-d$^{10}$ system are due to the hybrization of various Cu-d$^9$ configurations with the full O
band $2p^6$, in other words there is a single hole in the system and
the problem can be solved trivially. As expected, if the bottom of the oxygen band at $\Delta-4t_{pp}> \epsilon_d$, then the system lies in the positive charge-transfer regime, where the lowest energy electron removal state is
dominated by an (antibonding) orbital of $b_1$ symmetry that has
predominantly Cu-d$^9$ character; this is mixed with a ligand hole
$d^{10}\underline{L}$ states which have a low amplitude of
probability. This confirms that if there is a single hole in the
system, it is indeed located primarily on the Cu as in the ground state of the undoped cuprates. 

Photoemission or doping of the system with one hole from its ground state of mainly $d^9$ character removes another electron. The
resulting two-hole problem is exactly solvable using the Cini-Sawatzky
method\cite{CiniSawatzky}. The two-hole problem requires taking into
account the d-d Coulomb and exchange interactions $U(mm'm''m''')$
described by $V_{dd}$. These are listed in Table~\ref{interaction},
which contains the interaction matrices for all singlet/triplet
irreducible representations of the $D_{4h}$ point group spanned by two
$d$ holes, in terms of the Racah parameters $A, B$, and $C$. Throughout the paper, the free-ion values $B=0.15$ eV, $C=0.58$ eV are adopted and $A$ is treated as a variable.

\begin{table*}[t]
\caption{Irreducible representations spanned by two $d$ holes ($d^8$)
  and corresponding Coulomb and exchange matrix elements in terms of
  Racah parameters $A, B, C$. The basis functions are based on the
  single hole irreducible representations: $b_1(d_{x^2-y^{2}}),
  a_1(d_{3z^2-r^{2}}), b_2(d_{xy}), e_x(d_{xz}), e_y(d_{yz})$. Throughout the paper, the free-ion values $B=0.15$ eV, $C=0.58$ eV are adopted and $A$ as a variable is also often refereed to as Hubbard $U$, whose value varies in different materials.}
\centering
\begin{tabular*}{\textwidth}{@{\extracolsep{\fill} } ccccc}
 \hline\hline
 $\bf ^{1}A_1$ & $a^{2}_1$ & $b^{2}_1$ & $b^{2}_2$ & $(e^{2}_x+e^{2}_y)/\sqrt{2}$  \\ 
 \hline
 $a^{2}_1$ & $A+4B+3C$ & $4B+C$    & $4B+C$    & $\sqrt{2}(B+C)$     \\ 
 $b^{2}_1$ & $4B+C$    & $A+4B+3C$ & $C$       & $\sqrt{2}(3B+C)$    \\
 $b^{2}_2$ & $4B+C$    & $C$       & $A+4B+3C$ & $\sqrt{2}(3B+C)$   \\
 $(e^{2}_x+e^{2}_y)/\sqrt{2}$ & $\sqrt{2}(B+C)$    & $\sqrt{2}(3B+C)$   & $\sqrt{2}(3B+C)$   & $A+7B+4C$  \\
  &  & &  &     \\
\end{tabular*}
\\
\centering
\begin{tabularx}{\textwidth}{@{\extracolsep{\fill} } cccccc}
  $\bf ^{1}A_2$ & $b_1 b_2$ & $\bf ^{3}B_1$ & $a_1 b_1$ & $\bf ^{3}B_2$ & $a_1 b_2$  \\ 
 \hline
 $b_1 b_2$ & $A+4B+2C$ & $a_1 b_1$    & $A-8B$    & $a_1 b_2$    & $A-8B$  \\ 
  &  & &  & &    \\
\end{tabularx}
\\
\centering
\begin{tabularx}{\textwidth}{@{\extracolsep{\fill} } ccccccccc}
  $\bf ^{3}A_2$ & $b_1 b_2$ & $e_x e_y$ & $\bf ^{1}B_1$ & $a_1 b_1$ & $(e^{2}_x-e^{2}_y)/\sqrt{2}$ & $\bf ^{1}B_2$ & $a_1 b_2$ & $e_x e_y$  \\ 
 \hline
 $b_1 b_2$ & $A+4B$ & $6B$ & $a_1 b_1$  & $A+2C$  & $2\sqrt{3}B$ & $a_1 b_2$ & $A+2C$ & $2\sqrt{3}B$ \\ 
 $e_x e_y$ & $6B$ & $A-5B$ & $(e^{2}_x-e^{2}_y)/\sqrt{2}$   & $2\sqrt{3}B$  & $A+B+2C$  & $e_x e_y$ & $2\sqrt{3}B$ & $A+B+2C$ \\
  &  & &  & &    \\
\end{tabularx}
\\
\centering
\begin{tabularx}{\textwidth}{@{\extracolsep{\fill} } cccccccc}
  $\bf ^{1}E$ & $e_xb_1$ & $e_xa_1$ & $e_yb_2$ & $\bf ^{1}E$ & $e_yb_1$ & $e_ya_1$ & $e_xb_2$  \\ 
 \hline
 $e_xb_1$ & $A+B+2C$ & $-\sqrt{3}B$ & $-3B$         & $e_yb_1$ & $A+B+2C$ & $\sqrt{3}B$ & $3B$ \\ 
 $e_xa_1$ & $-\sqrt{3}B$ & $A+3B+2C$ & $-\sqrt{3}B$ & $e_ya_1$ & $\sqrt{3}B$ & $A+3B+2C$ & $-\sqrt{3}B$  \\
 $e_yb_2$ & $-3B$ & $-\sqrt{3}B$ & $A+B+2C$         & $e_xb_2$ & $3B$ & $-\sqrt{3}B$ & $A+B+2C$  \\\\   
\end{tabularx}
\\
\centering
\begin{tabularx}{\textwidth}{@{\extracolsep{\fill} } cccccccc}
  $\bf ^{3}E$ & $e_xb_1$ & $e_xa_1$ & $e_yb_2$ & $\bf ^{3}E$ & $e_yb_1$ & $e_ya_1$ & $e_xb_2$  \\ 
 \hline
 $e_xb_1$ & $A-5B$ & $-3\sqrt{3}B$ & $3B$         & $e_yb_1$ & $A-5B$ & $3\sqrt{3}B$ & $-3B$ \\ 
 $e_xa_1$ & $-3\sqrt{3}B$ & $A+B$ & $-3\sqrt{3}B$ & $e_ya_1$ & $3\sqrt{3}B$ & $A+B$ & $-3\sqrt{3}B$  \\
 $e_yb_2$ & $3B$ & $-3\sqrt{3}B$ & $A-5B$         & $e_xb_2$ & $-3B$ & $-3\sqrt{3}B$ & $A-5B$  \\ 
\hline\hline 
\end{tabularx}
\label{interaction}
\end{table*}

If we chose to focus on relevance to experiments, we would need to
calculate the $d$-electron removal spectrum $A^{\Gamma}_d(\omega)$ which can be
compared to photoemission experiments, and the $d^8$ partial density
of states (PDOS) for the various two-hole irreducible representations
(symmetry channels) $A^{\Gamma}_{d^8}(\omega)$, linked to the resonant
photoemission.\cite{Eskes90} They are defined by
\begin{align} \label{rho}
A^{\Gamma}_d(\omega) &= -\frac{1}{\pi} \sum_{mm'} \lim_{\delta\rightarrow 0} \Im G_{dd}(m,m',\omega+i\delta;\Gamma)
\nonumber \\ A^{\Gamma}_{d^8}(\omega) &= -\frac{1}{\pi} \sum_{mm'} \lim_{\delta\rightarrow 0} \Im
G_{d^8}(m,m',\omega+i\delta;\Gamma)
\end{align}
with
\begin{align} \label{Gs}
G_{dd}(m,m',z;\Gamma) &= \langle \psi_{g.s.}| d^{\phantom\dagger}_{m'} \hat{G}(z)
d^{\dagger}_{m} | \psi_{g.s.} \rangle \nonumber \\ G_{d^8}(m,m',z;\Gamma) &= \langle 0|
d^{\phantom\dagger}_{m'} d^{\phantom\dagger}_{m} \hat{G}(z) d^{\dagger}_{m} d^{\dagger}_{m'}
|0 \rangle \nonumber \\ \hat{G}(z) = (z-& \hat{H})^{-1}, z=\omega+i\delta
\end{align}
Here, $|0\rangle$ is the Cu-3d$^{10}+ $ O-2p$^6$ state, {\em i.e.} the state
with no holes, while $| \psi_{g.s.}\rangle$ is the one-hole ground-state.

For our purposes, however, it suffices to obtain their common
part, namely the component of $d^8$ partial density of states
$A^{\Gamma}(\omega)$ which assumes that one hole has already occupied the $b_1$
orbital (remember that $d^{\dagger}_{b_1} |0 \rangle$ is the dominant contribution
to $| \psi_{g.s.} \rangle$):
\begin{align} \label{Aw}
& A^{\Gamma}(\omega) = -\frac{1}{\pi} \sum_{m} \lim_{\delta\rightarrow 0} \Im G_{d}(m,b_1,\omega+i\delta;\Gamma)
  \nonumber \\ & G_{d}(m,z;\Gamma) = \langle 0| d^{\phantom\dagger}_{b_1}
  d^{\phantom\dagger}_{m} \hat{G}(z) d^{\dagger}_{m} d^{\dagger}_{b_1} |0 \rangle
\end{align}
We focus primarily on $G_{d}(m,z;\Gamma)$ from now on, but all
other propagators $G_{dd}(m,m',z;\Gamma)$ and $G_{d^8}(m,m',z;\Gamma)$ can be
calculated similarly.

\subsection{Variational exact diagonalization}

We use variational exact diagonalization to calculate the propagator
$G_{d}(m,z;\Gamma)$. The two-hole states in the variational space are of
three possibile types: (a) both holes are on the Cu; (b) one hole is
on the Cu and one on an O; and (c) both holes are on O sites. All
states in (a) are included in the variational space. For the (b) and
(c) states, we impose a cutoff ${\bf R_c}$ between the O hosting the
hole(s) and the Cu. Obviously, $\bf R_c \rightarrow \infty$ recovers the full Hilbert
space. We typically set $R_c=20$ for the results shown below. 
{This
suffices for convergence to be reached for all the bound states. Unless we use a very large $\eta$, the continua are not yet fully converged for this $R_c$, instead they look like a collection of peaks whose number increases with $R_c$. The upper and lower bandedges are already converged, however, and that is all the information relevant for our analysis.}

\color{black}{
Within this variational space, we set up the Hamiltonian matrix for
each irreducible representation and use standard exact diagonalization
to calculate the corresponding propagators via Lanczos
diagonalization.}

\section{Results}\label{results}
\begin{figure*} 
\psfig{figure=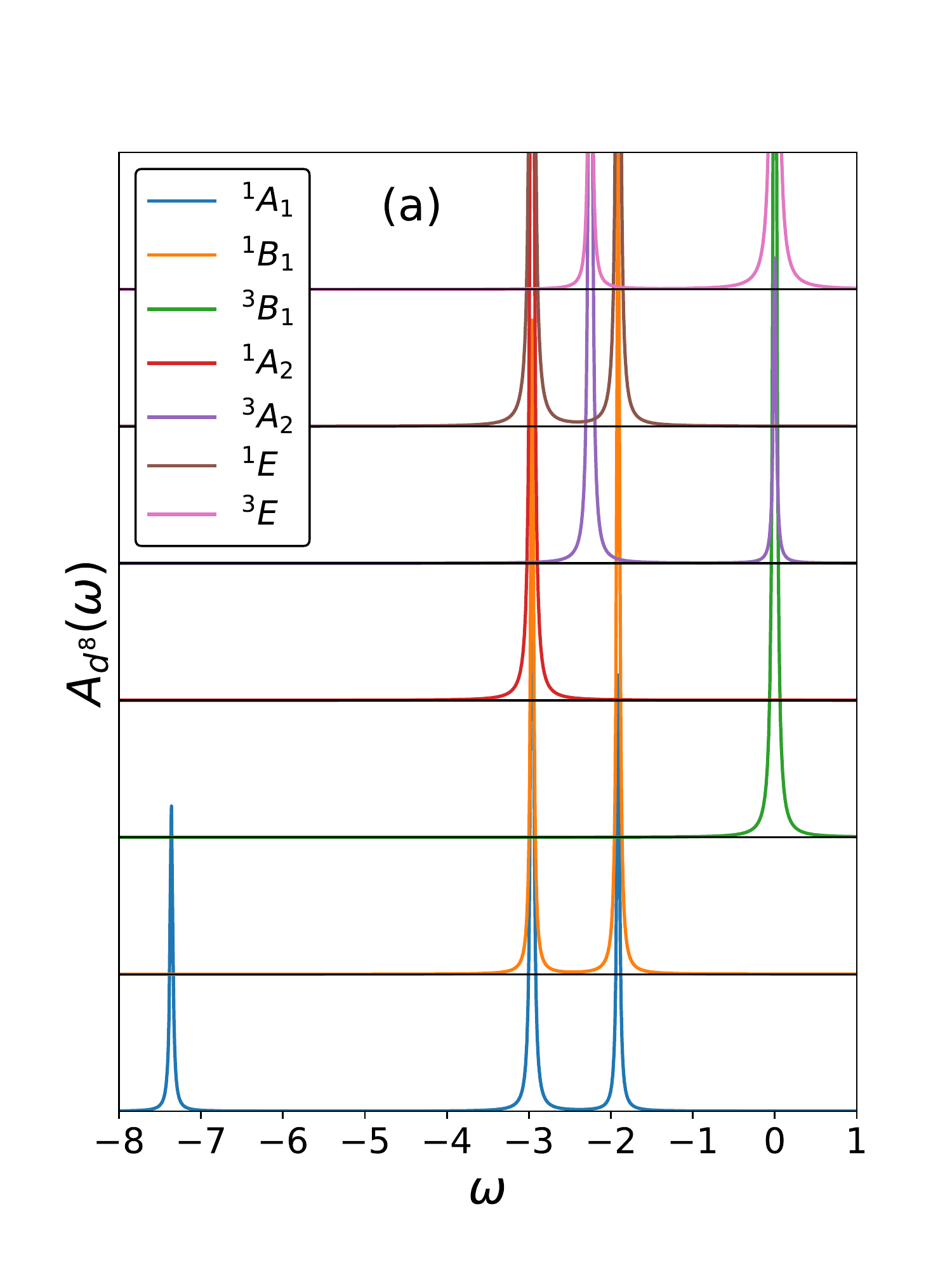}.pdf,
height=5.0cm,width=4.0cm,angle=0,clip}
\psfig{figure=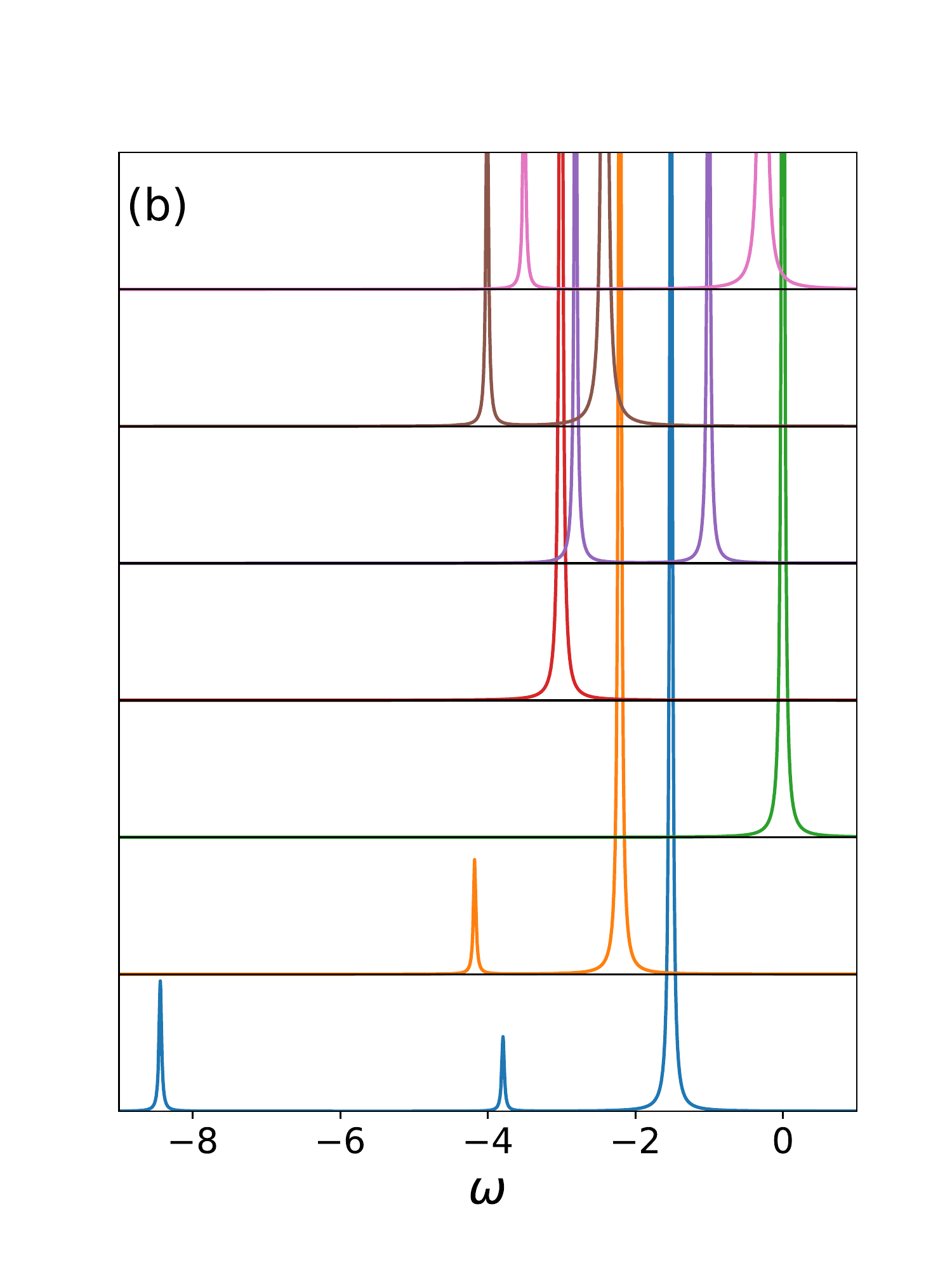}.pdf,
height=5.0cm,width=4.0cm,angle=0,clip}
\psfig{figure=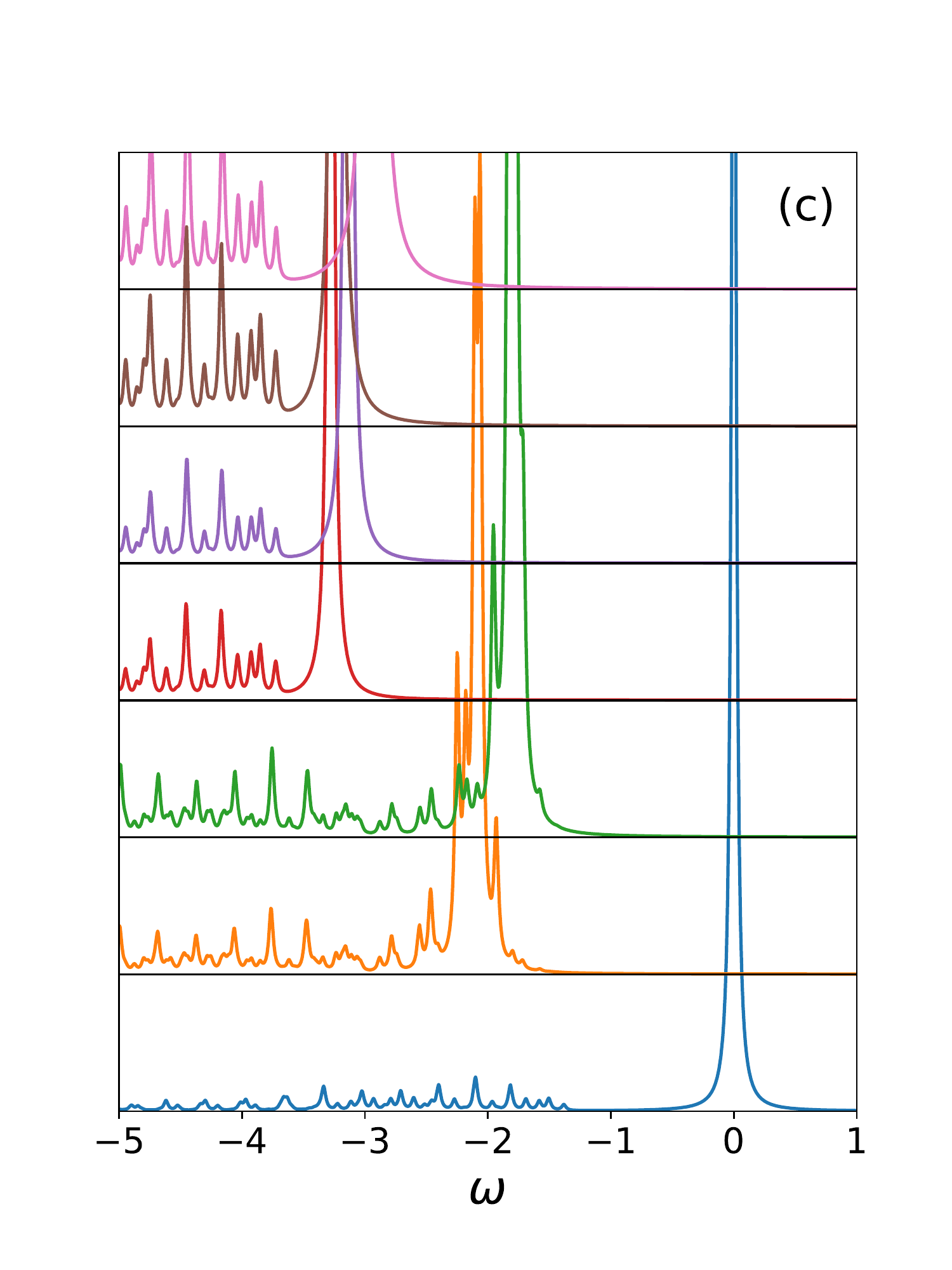}.pdf,
height=5.0cm,width=4.0cm,angle=0,clip} 
\psfig{figure=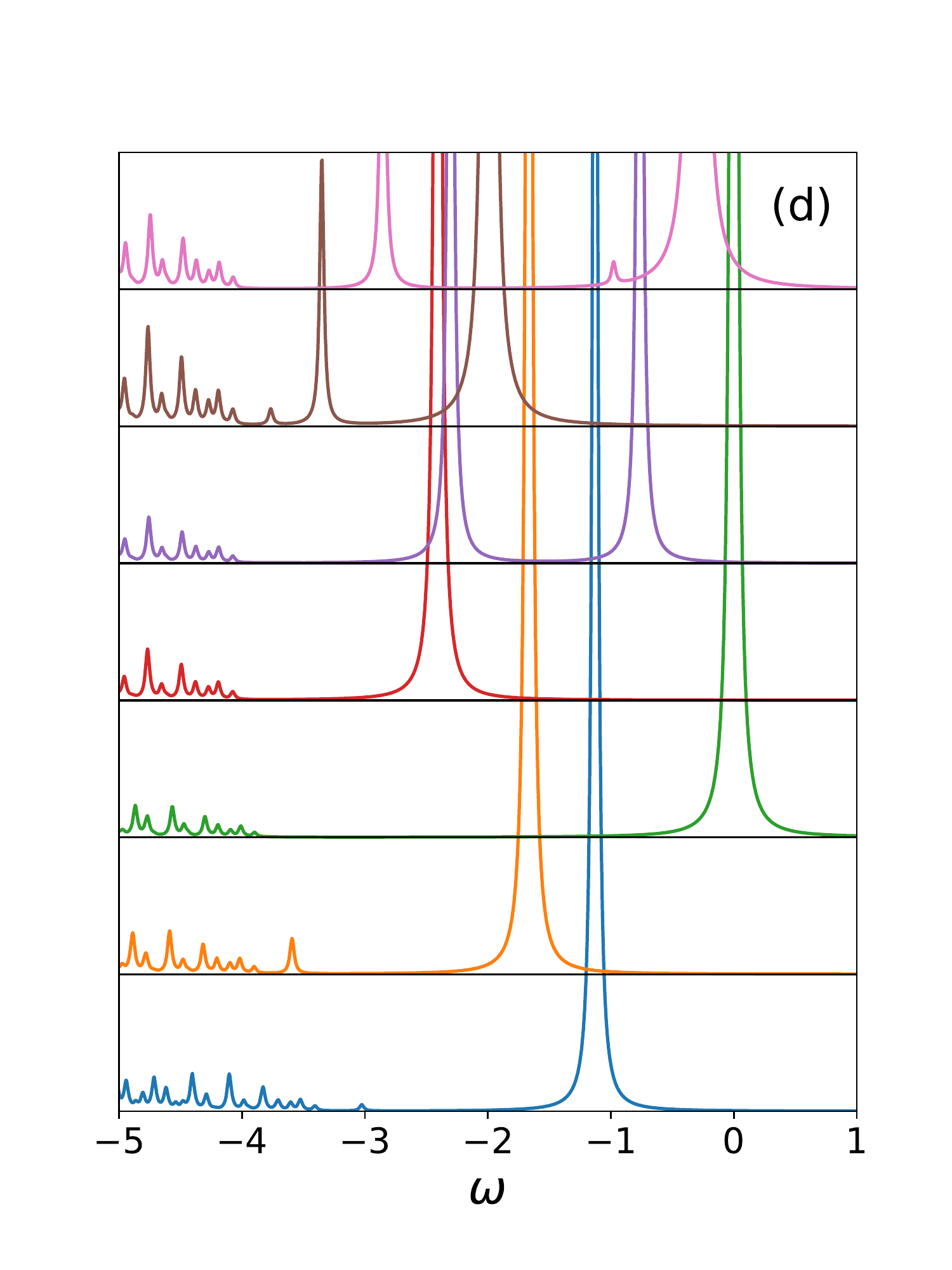}.pdf,
height=5.0cm,width=4.0cm,angle=0,clip}
\caption{(Color online) The two-hole spectra $A^{\Gamma}(\omega)$ calculated for various irreducible representations $\Gamma$ in the seven-orbital (N7) model in the cases of (a) single Cu-$d^8$ ion with onsite energies $\epsilon_d(m)=0$, (b) single Cu-$d^8$ ion including the additional ligand field splitting~\cite{ligandfield}, and single Cu impurity in a lattice of oxygen with finite Cu-O hybridization corresponding to two characteristic cases with differing two-hole ground state of $^1\!A_1$ (c) and $^3\!B_1$ (d) symmetry respectively.  
The parameters are (c) $\Delta=2.75$ eV, $A=6.5$ eV and (d) $\Delta=6.5$ eV, $A=2.5$ eV with $t_{pd}=1.5$ eV, $t_{pp}=0.55$ eV. The chemical potential, taken to be zero energy, is chosen as the lowest energy of two-hole state.}
\label{atomic}
\end{figure*}

Before proceeding, we remark that throughout the paper, we adopt the usual convention of photoemission spectroscopies that the electron removal energy, or the hole energy,  increases to the left while the energy of the electron addition states increases to the right. The chemical potential, taken to be the  zero energy, is chosen at the lowest energy of the two-hole state.

Figure~\ref{atomic} illustrates the two-hole spectra $A^{\Gamma}(\omega)$ for various irreducible representations $\Gamma$ in the seven-orbital (N7) model in the cases of (a) single Cu-$d^8$ ion with onsite energies $\epsilon_d(m)=0$, (b) single Cu-$d^8$ ion including the additional ligand field splitting~\cite{ligandfield}, and single Cu impurity in a lattice of oxygen with finite Cu-O hybridization corresponding to two characteristic cases with differing two-hole ground states of $^1\!A_1$ (c) and $^3\!B_1$ (d) symmetry respectively. 
In the limiting case of $t_{pd}=0$, the two $d$ holes can have $^{1}\!S, ^{3}\!P, ^{1}\!D, ^{3}\!F, ^{1}\!G$ configurations, whose energies are listed {\em e.g.} in  Ballhausen.\cite{Ballhausen}
As shown in Fig.~\ref{atomic}(a), our two-hole spectra
$A^{\Gamma}_{d^8}(\omega)$ indeed consist of one or more discrete peaks located
at these energies; the number of peaks and their corresponding spectral weights depend on the singlet/triplet nature of the irreducible representation $\Gamma$.

Note that the inclusion of the ligand field splittings in Fig.~\ref{atomic}(b) only induces modest shifts of the peaks and modification of their spectral weights. In contrary, hybridization with the O band results in a significant spreading of the spectral weights over a much wider energy range, and a complete re-ordering of the low-energy, multiplet-like bound states. Indeed, for this realistic value of $t_{pd} = 1.5$ eV, there is no significant correspondence between the bound peak positions and the multiplets in the atomic limit of $t_{pd} = 0$; neither the splittings between the bound peaks, nor even their order, mimic what is found in the atomic multiplet.  Instead, of great importance is that for a not too large $\Delta$ (see panel c), the lowest energy state is not the expected triplet according to the Hund's rule but a singlet state; and the first triplet state lies at more than 1.5 eV higher energy. Therefore, our results caution against the approach of using Wannier functions together with $A, B, C$ Racah parameters renormalized so as to obtain an atomic limit multiplet similar to the one produced by the strong hybridization. Fig.~\ref{atomic} demonstrates that the two have very different splittings and even ordering of the peaks in the various symmetry channels. 

From now on we focus on the case where the lowest energy two-hole ground state is of $^1\!A_1$ symmetry, as illustrated in Fig.~\ref{atomic}(c).
There is a clearly visible low-energy discrete peak, proving that for these parameters the two holes form a bound state which is a linear combination of two holes on Cu, one hole on Cu and the other on O, and two holes on O, i.e. the configurations of $d^8$, $d^9L$ and $d^{10}L^2$. The lowest energy state for adding one hole to a Cu $d^9$ state as in the hole doped cuprates would be a bound state of $^1\!A_1$ symmetry similar to the ZRS as the lowest energy state. According to the Fig.~\ref{atomic}, this bound state would be separated from a continuum correponding to the doped hole in an O-$2p$ band by about 1 eV, which is indeed close to what is observed in ARPES experiments of the cuprates. In addition to the broadening and appearance of bound states beyond the continua, the hybridization also introduces the ligand field like splittings which will mix the various atomic multiplets.
Note that for these parameter values, only the $^1\!A_1$ peak is clearly below the correponding continuum, and thus a truly bound state; the other peaks are inside the lower edge of their continua. At even higher energies lies the two hole continuum, where both holes move
freely in the O lattice and the Cu is in a $d^{10}$ state; this is superimposed over strong resonances where Cu multiplet lines hybridize with (and are shifted around by) this
continuum. All this forms a very broad structure with mixed character and is basically the origin of the so called ``waterfall'', a {name coined} by Lanzara et al~\cite{Lanzara}.

Of most interest are three lowest peaks, of which the lowest one, with $^1\!A_1$ symmetry, is the first ionization state starting from Cu-d$^9$. Its eigenstate is
\begin{align} \label{eq:HM}
  |\psi \rangle = &\sqrt{0.072} |b_1b_1 \rangle + \sqrt{0.549} |b_1 L_{b_1} \rangle + \sqrt{0.054} |b_1 L^{'}_{b_1} \rangle
  \nonumber\\ & + \sqrt{0.275} |d^{10}L^{2} \rangle + \dots 
\end{align}
where $\dots$ represents states having $a_1a_1, b_2b_2, ee$ characters,
whose probabilities add up to less than $1\%$. Here $L_{b_1}$ denotes
one hole in a linear combination of O orbitals nearest to the Cu impurity, with overall $b_1$
symmetry. We emphasize that this weight distribution is almost independent on the number of orbitals considered, whether the N3, N7, N9, or N11 models. This shows that the ground-state is only about 55\% ZRS-like, {\em i.e.} $|b_1 L_{b_1} \rangle$. $L^{'}_{b_1}$ denotes the configurations where the hole is on the second, third, {\em etc.} rings of O ions, which strictly speaking are
discarded by the ZRS. The strong mixing of the ground state with the $d^{10}L^2$ state is the reason for the strong antiferromagnetic exchange interaction which stabilizes the singlet. It is worth noting that this strong wave function mixing is {strongly dependent} on $t_{pd}$, which in turn is strongly dependent on the interatomic distance between Cu and the nearest-neighbor O. This is the origin of a {possibly} strong electron-phonon and magnon-phonon coupling. 
 
The second and third lowest peaks are the high spin $^3\!B_1$ state and the singlet $^1\!B_1$ state respectively. 
All these results are qualitatively similar to those reported in
previous work by Eskes {\em et al}.\cite{Eskes88,Eskes90}.
The quantitative differences, especially the differences in the weights of various continua, are due
to how the O band is modelled (realistic tight-binding model in our
work, {\em vs.} featureless semi-elliptical DOS in theirs). 

Next we elaborate on the case where the lowest energy two-hole ground state is of $^3\!B_1$ symmetry, as illustrated in Fig.~\ref{atomic}(d). To obtain this we adopted $A-\Delta=-4$ eV, which puts the system well into the Mott Hubbard rather than charge transfer gap of the ZSA clasification scheme. The major difference from the case shown in Fig.~\ref{atomic}(c) is the order of lowest peaks, which changed to be of $^3\!B_1, ^3\!E, ^3\!A_2$ symmetries from those of $^1\!A_1, ^3\!B_1, ^1\!B_1$ symmetries. Furthermore, it is clear that the conventional three-orbital (N3) model cannot capture the lowest bound state any more due to the lack of the involvement of the $a_1(d_{3z^2-r^{2}})$ orbital.

\begin{figure*} 
\psfig{figure=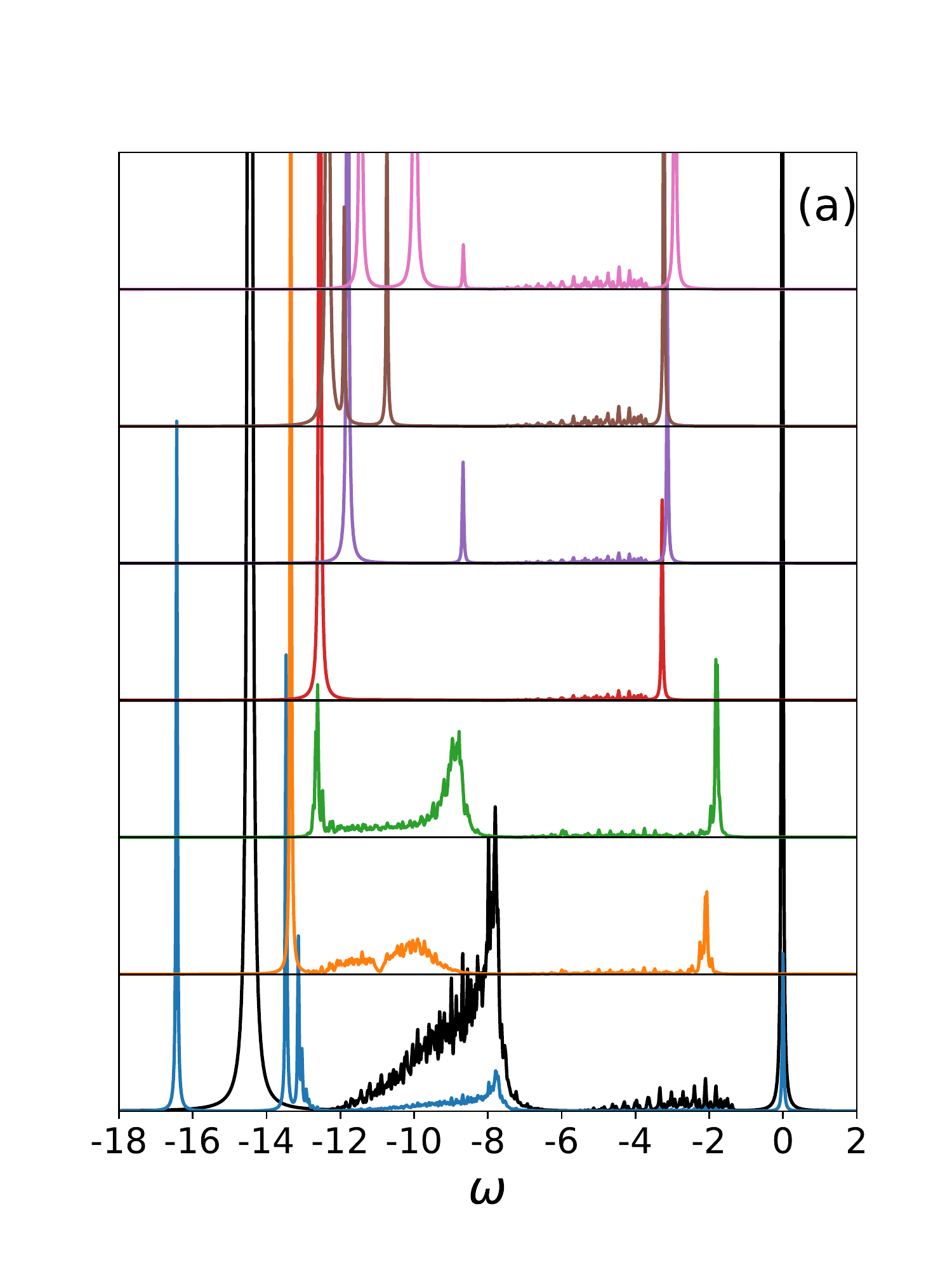}.pdf,
height=6.5cm,width=5.5cm,angle=0,clip}
\psfig{figure=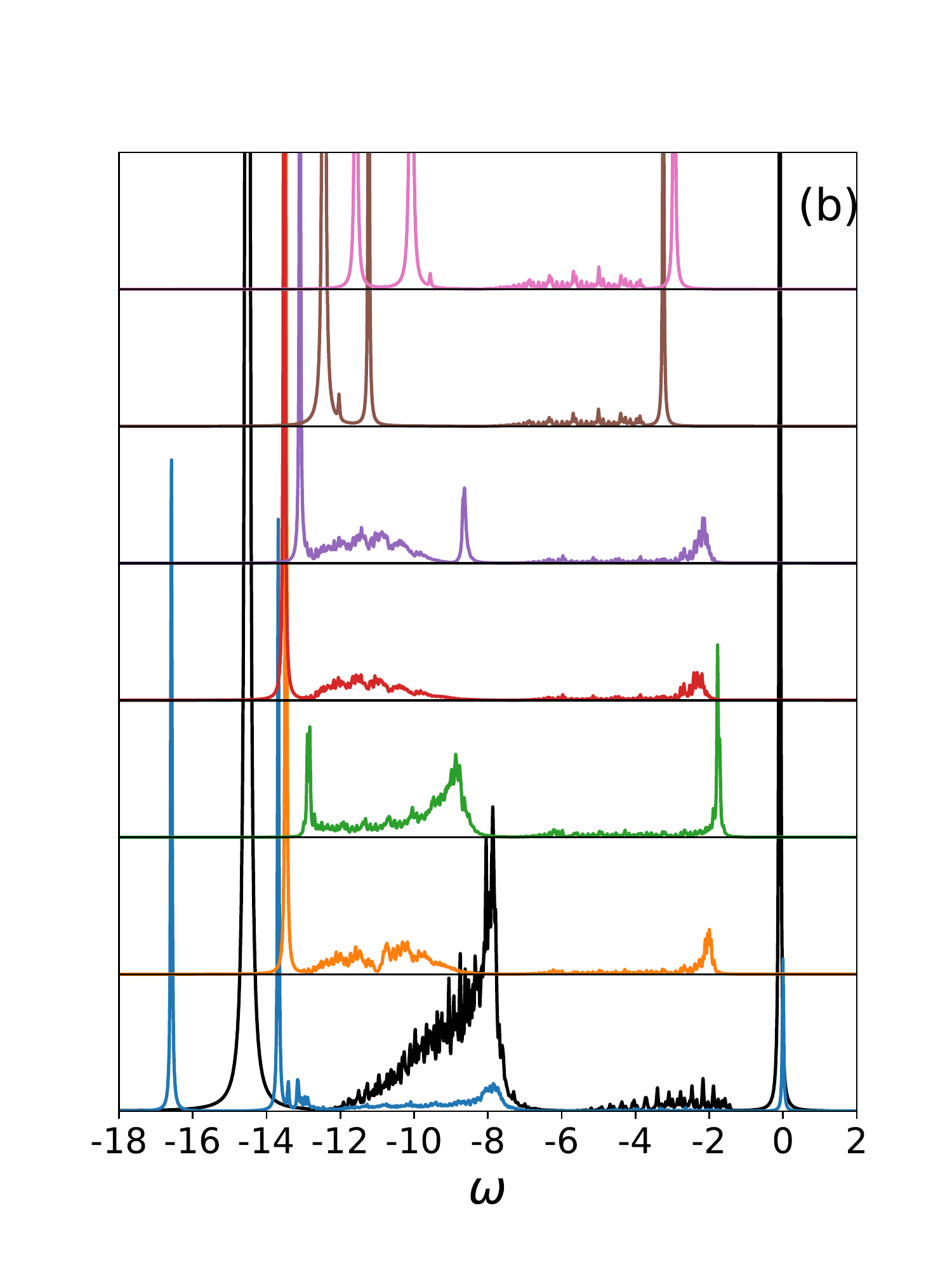}.pdf,
height=6.5cm,width=5.5cm,angle=0,clip}
\psfig{figure=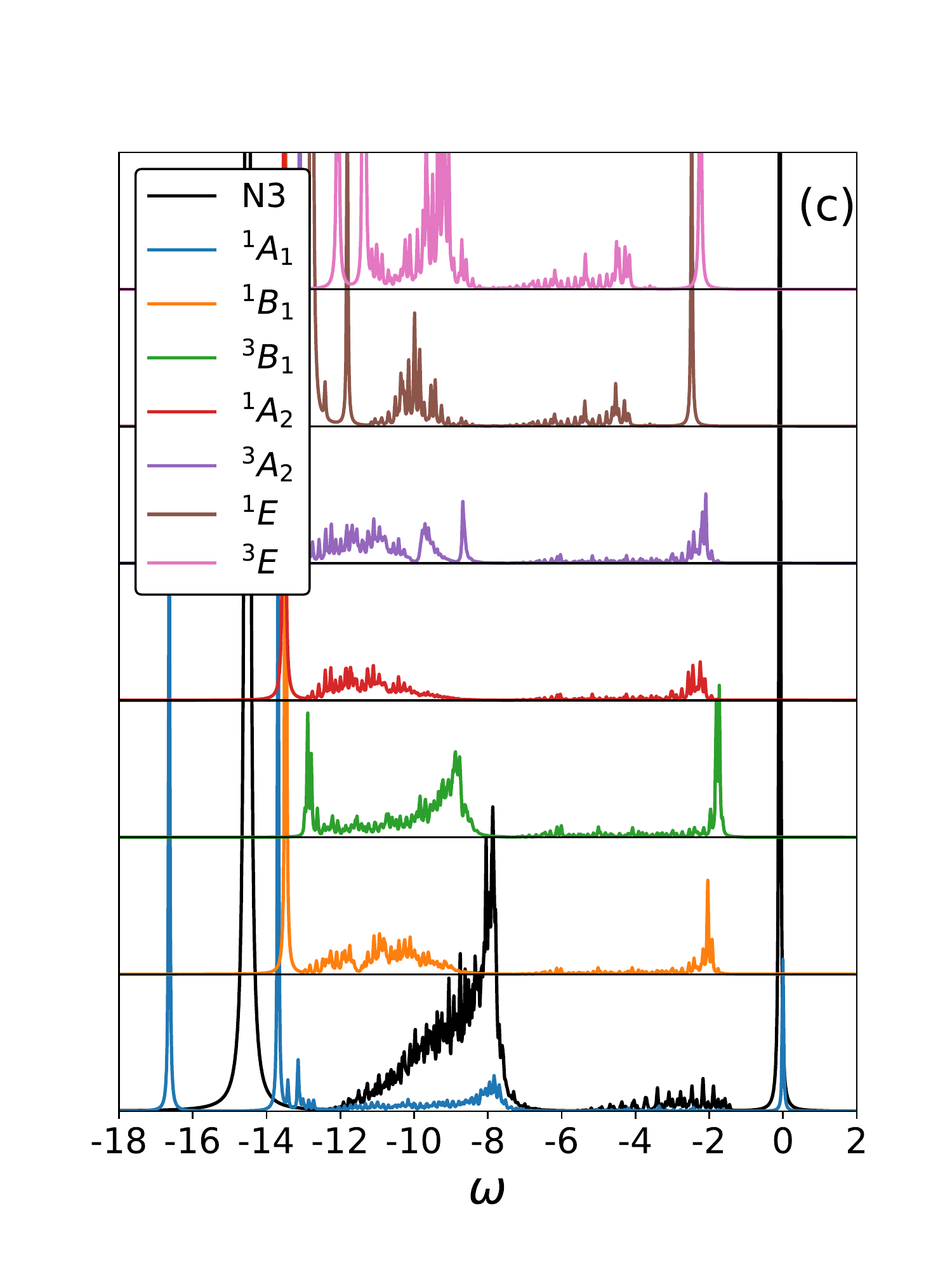}.pdf,
height=6.5cm,width=5.5cm,angle=0,clip} \\
\psfig{figure=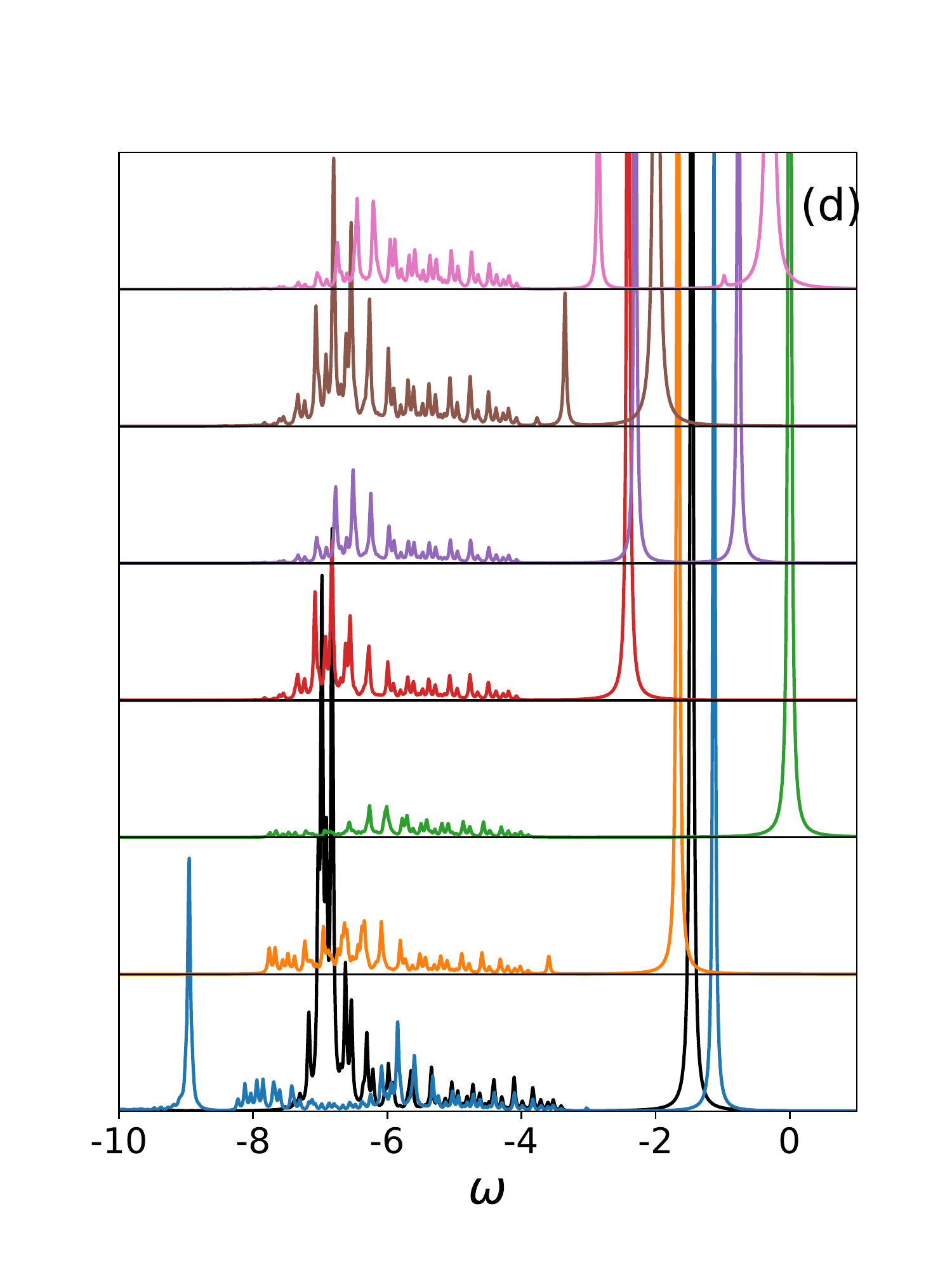}.pdf,
height=6.5cm,width=5.5cm,angle=0,clip}
\psfig{figure=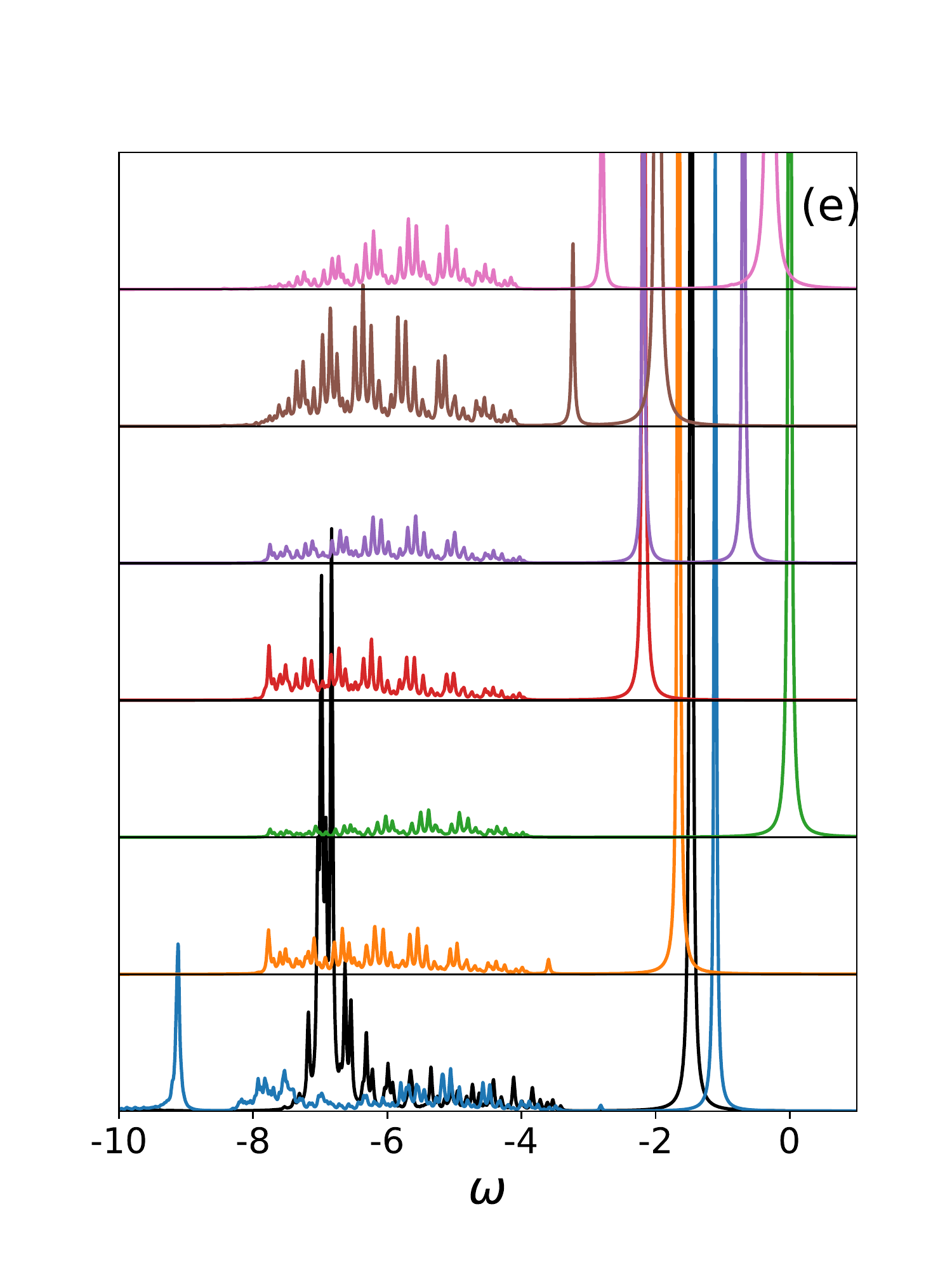}.pdf,
height=6.5cm,width=5.5cm,angle=0,clip}
\psfig{figure=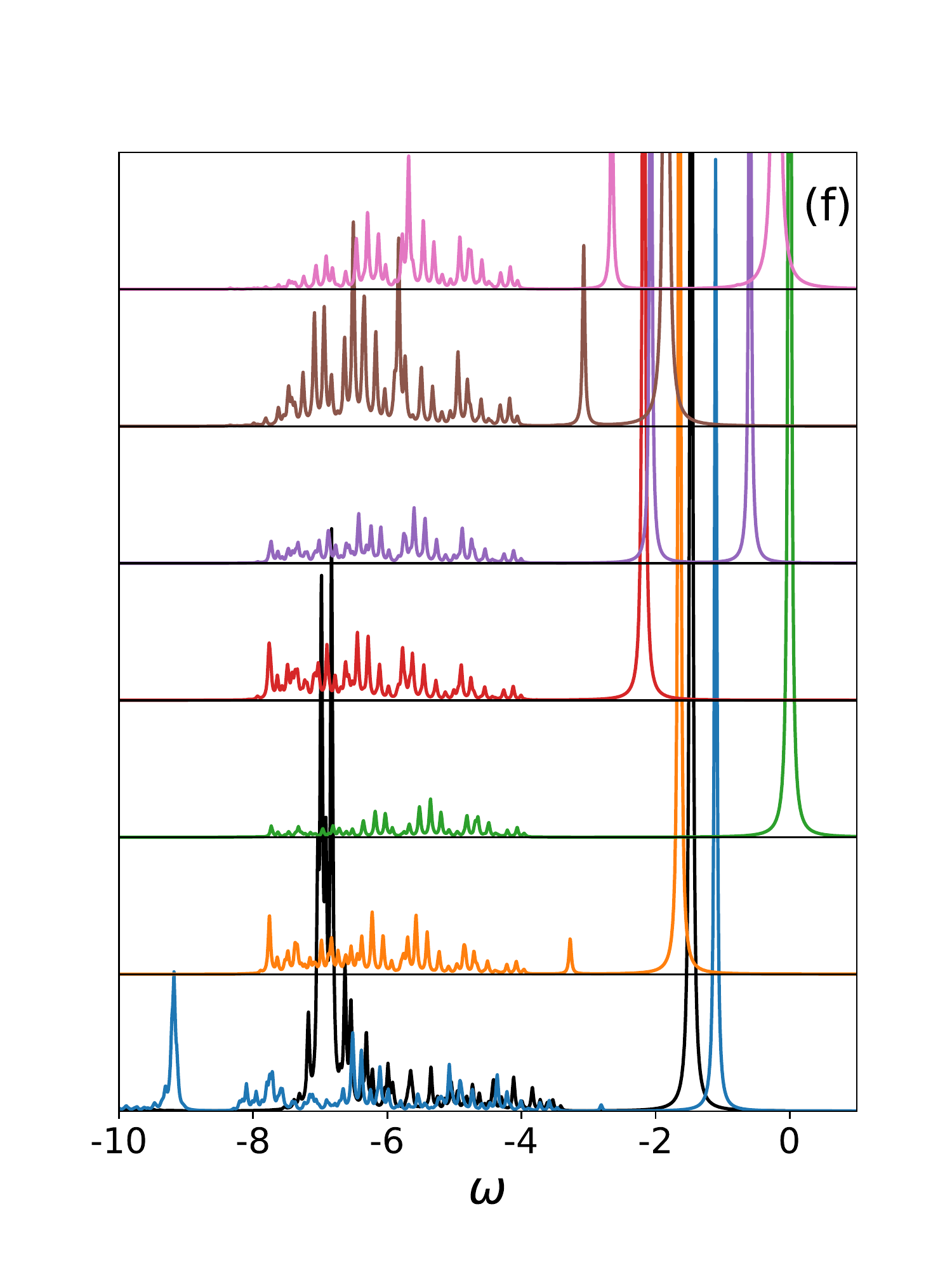}.pdf,
height=6.5cm,width=5.5cm,angle=0,clip}
\caption{(Color online) The comparison of two-hole spectra $A^{\Gamma}(\omega)$ calculated for various irreducible representations $\Gamma$ in the (a,d) seven-orbital (N7), (b,e) nine-orbital (N9), and (c,f) eleven-orbital (N11) models for two characteristic parameter sets corresponding to the low spin (singlet) (a-c) and high spin triplet (d-f) cases. The spectra of $N3$ model (black curve) is plotted for comparison as well.
The parameters are (a-c) $\Delta=2.75$ eV, $A=6.5$ eV and (d-f) $\Delta=6.5$ eV, $A=2.5$ eV with (a, d) $t_{pd}=1.5$ eV, $t_{pp}=0.55$ eV and (b-c, e-f) $t_{pd\sigma}=\sqrt{3}$ eV, $t_{pd\pi}=0.75$ eV, $t_{pp\sigma}=0.9$ eV, $t_{pp\pi}=0.2$ eV. For the N3 model, we use $U_{dd}=A+4B+3C$, $t_{pp}=0.55$ eV.}
\label{spectra_imp}
\end{figure*}

To investigate the effects of including more Cu-3d and/or O-2p orbitals in the model Hamiltonians, Figure~\ref{spectra_imp} compares the two-hole spectra $A^{\Gamma}(\omega)$ calculated for
the (a,d) seven-orbital (N7), (b,e) nine-orbital (N9), and (c,f) eleven-orbital (N11) models for two characteristic parameter sets corresponding to the low spin (singlet) (a-c) and high spin triplet (d-f) cases. 
The comparison between N7 and N9/N11 models illustrates the impact of including additional $\pi$-bonding oxygen orbitals. The additional hybridization with Cu-$b_2(d_{xy})$ orbital
extends the continua to lower energies for all the symmetries, which causes a much smaller difference between the continuum bottom of various A and B types of symmetries. This clearly demonstrates the importance of having all the continua in place correctly in order to decide which is the lowest energy state. For example, if the $^1\!A_1$ continuum would also be involved in the hybridization with the $^3\!B_1$ or $^1\!B_1$ state, these states would be appreciably closer to the $^1\!A_1$ lowest energy state and even cross it. This could happen if we could take into account the full lattice of Cu-$d^9$ states in the starting configuration, for example, as done in the exact diagonalization study of the large cluster with 32 Cu sites and 64 O sites by Lau et al~\cite{Lau}. It is important to note that Lau indeed found a very strong ferromagnetic coupling between the Cu sandwiching an O hole, which indicates that our impurity limit could be different from what happens in the actual crystal although the experiments of cuprates did agree with our classification for the undoped system. Strong hole doping howeve could strongly modify these conclusions. This also  questions the use of single site DMFT or single orbital cluster DMFT results with regard to the relevance for the full problem which includes both O and Cu states explictly in the cluster.  

\begin{figure}[!ht]
\psfig{figure=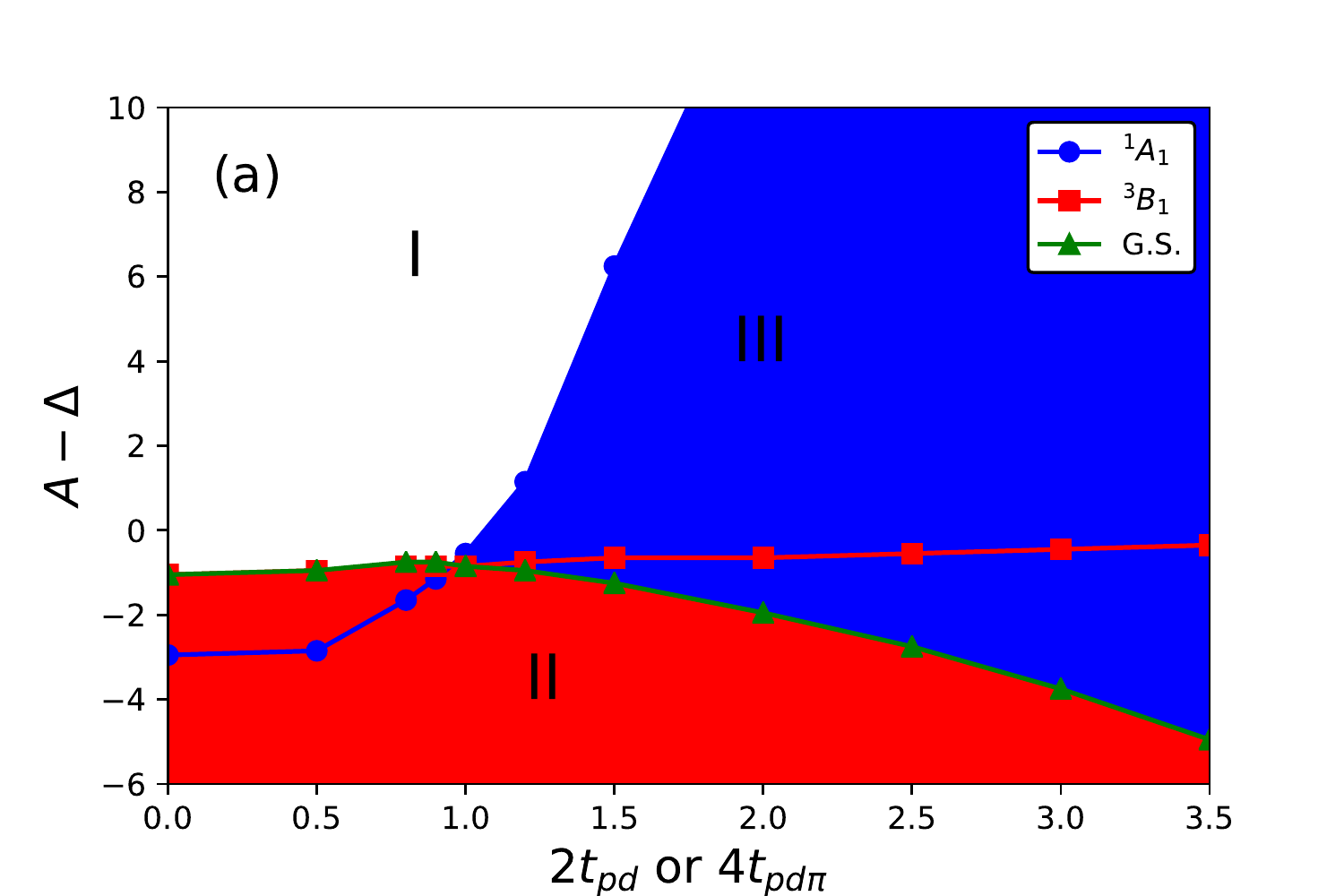}.pdf,height=5.5cm,width=8.0cm,angle=0,clip}
\psfig{figure=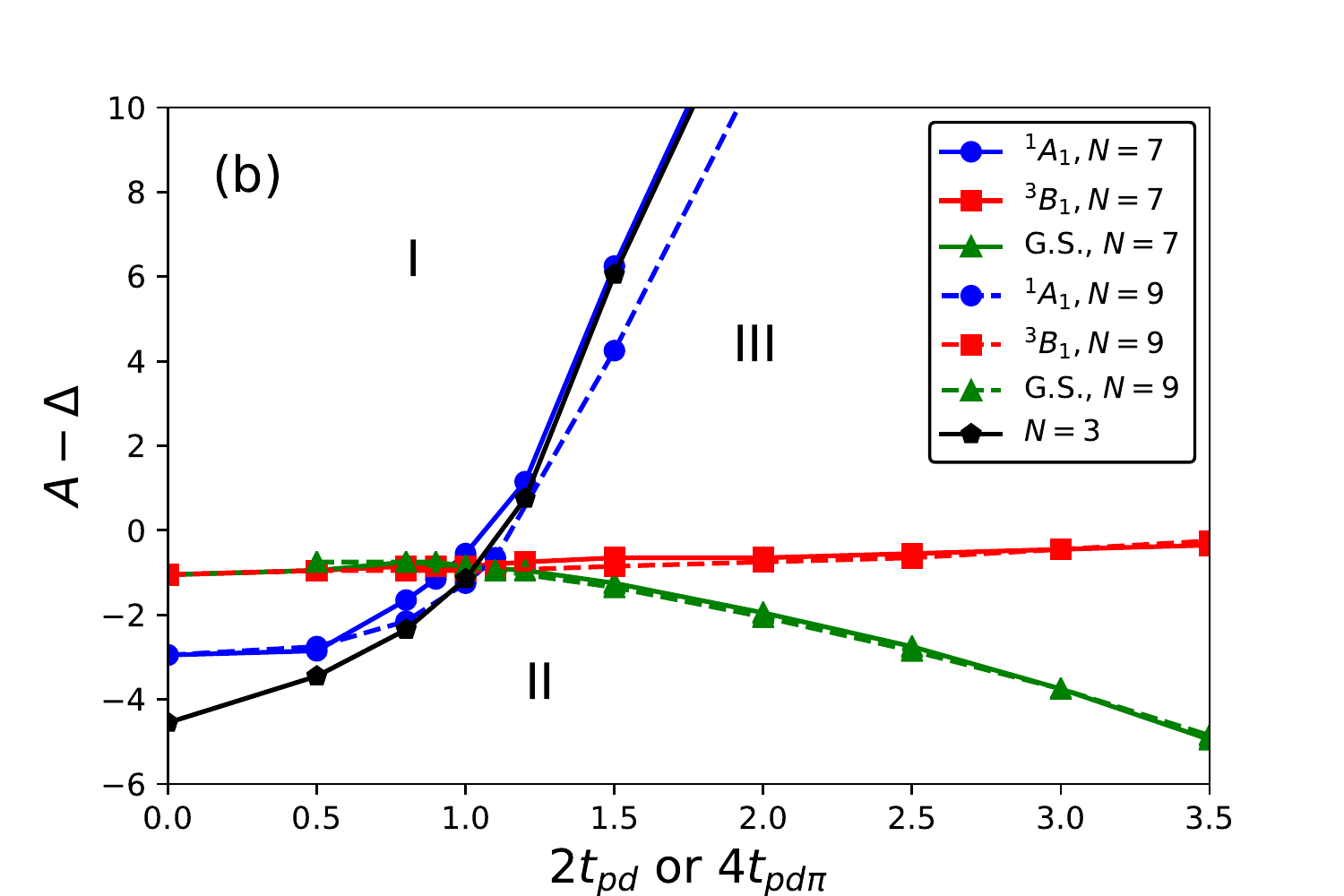}.pdf,height=5.5cm,width=8.0cm,angle=0,clip}
\caption{(Color online) (a) N7 one-doped hole phase diagram for
  $\Delta=2.75$ eV and oxygen bandwidth $W=4.4$ eV, {\em i.e.} $t_{pp}=0.55$ eV.
  Region I has no bound state, while in regions II and III and the
  doped hole is bound to the Cu hole in a complex with $^{3}\!B_1$ and
  $^{1}\!A_1$ symmetry, respectively. (b) Comparison between N3, N7
  and N9 phase diagrams. The conventional relations $t_{pd} \approx \sqrt{3}
  t_{pd\sigma}/2 = 2t_{pd\pi}$ and $t_{pp\sigma}=0.9$ eV, $t_{pp\pi}=0.2$ eV are adopted in the N9 model.
  For the N3 model, we use $U_{dd}=A+4B+3C$, $t_{pp}=0.55$ eV. The black
  line denotes the boundary for the appearance of the ZRS like states in the N3 model. The colored lines indicate the phase boundaries for obtaining a sharp ``bound like state'' at low energy with $^{1}\!A_1$ (blue curve) and $^{3}\!B_1$ (red curve) symmetries.}
\label{phase_diagram}
\end{figure}

\begin{figure}[t]
\psfig{figure=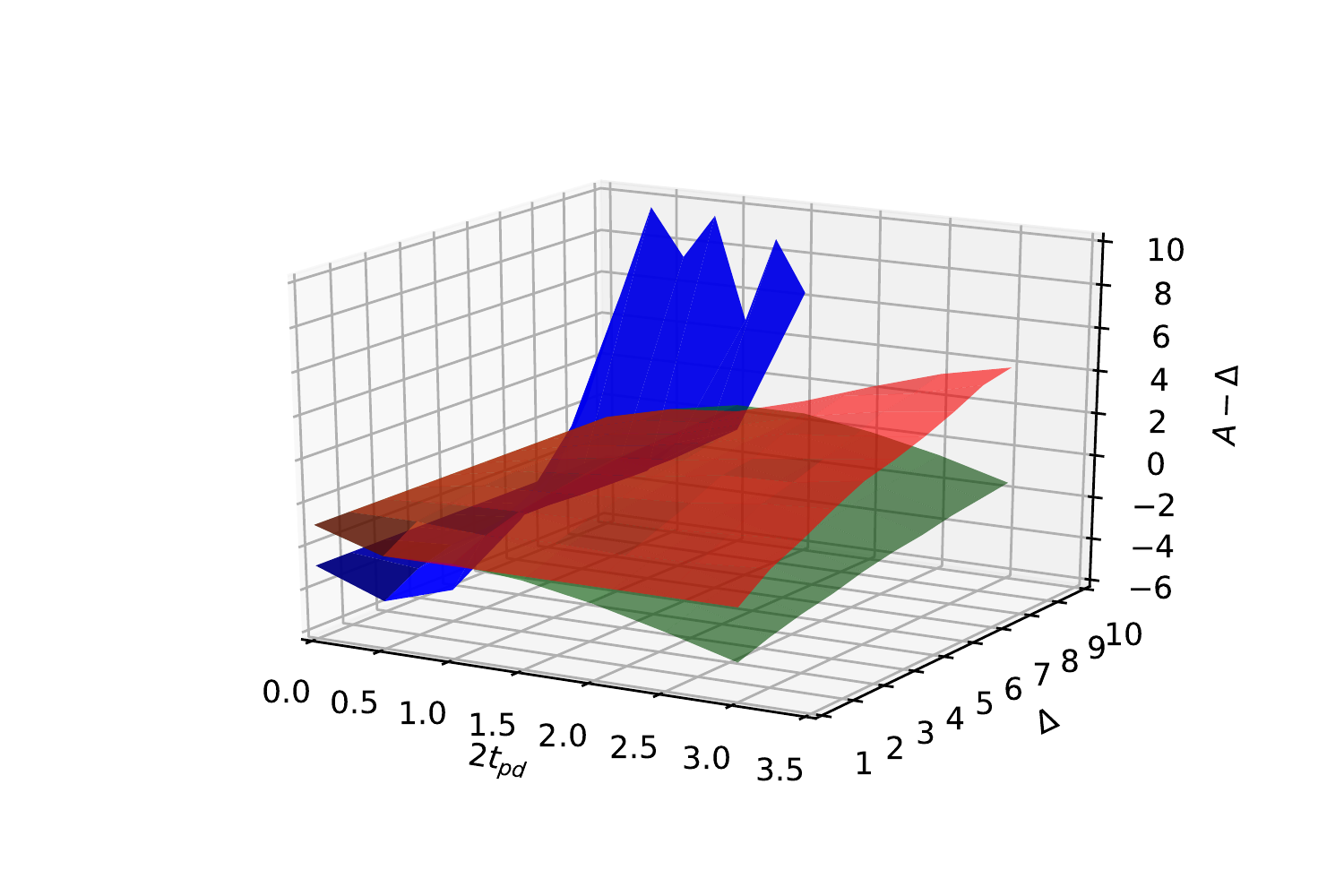}.pdf,height=5.0cm,width=7.0cm,angle=0,clip=true,
  trim = 2.5cm 1.0cm 0.5cm 1.5cm}
\caption{(Color online) Weak dependence of the phase boundaries of N7 phase diagram showed in
  Figure~\ref{phase_diagram}(a) (with the same color conventions) on
  the charge transfer energy $\Delta$.}
\label{phase_diagram3d}
\end{figure}

In the isolated Cu atom, the two-hole ground-state has $^3\!B_1$
symmetry (Hund's rule), while, as shown in Fig.~\ref{atomic}(c),
a strong enough hybridization with the O bands favors a ground-state
with $^1\!A_1$ symmetry, i.e. there is a high spin to low spin transition. To fully characterize the various possible symmetries of the ground-state, in Figure~\ref{phase_diagram} we show
phase diagrams in the full parameter space. In panel (a), we plot a
$A-\Delta$ {\em vs.} $2t_{pd}$ phase diagram, which can be directly
compared against that shown in Ref. \onlinecite{Eskes88}. It shows the
phase boundaries for obtaining the lowest peak with $^{1}\!A_1$ (blue
curve) and with $^{3}\!B_1$ (red curve) symmetries, respectively, for
an O bandwidth $W=4.4$ eV ($t_{pp}=0.55$ eV). Furthermore, the green curve
shows the phase boundary separating the ground state of $^{1}\!A_1$
(low spin) and $^{3}\!B_1$ (high spin) character. The three different
types of ground-states are filled by different colors: region I
denotes the absence of a bound ground-state state, {\em i.e.} the
doped hole moves freely in the O lattice instead of being bound to the
Cu hole. In regions II and III there is a bound ground state with
$^{3}\!B_1$ and $^{1}\!A_1$ symmetry, respectively. Clearly, region
III is physically relevant to cuprates.

While this phase diagram is  qualitatively similar with Eskes's
corresponding phase diagram,\cite{Eskes88} there are again
quantitative differences between the two. There is a shift of the
critical value of the $pd$ hybridization needed to obtain a bound
state with $^{1}\!A_1$ symmetry from their value
$T(B_{1g})=2t_{pd}\approx1.6$ eV to our value of $ \approx 1.0$ eV. In addition, the
lines separating the various regions have quite different slopes.
These non-trivial quantitative differences are  due to the
difference in how the O bath is modeled. One of the main reasons for this difference is that in the Eskes approach the ligand hole states are all spread equally over the hemispherical band while in our tight-binding band structure the $b_1$ symmetry hole states are concentrated at the bottom of the hole density of states making the appearance of a two-hole $^{1}\!A_1$ bound state possible at even lower $t_{pd}$. Another important difference caused by the same effect is that in our case the splitting between the $^{1}\!A_1$ and the $^{3}\!B_1$ peaks is larger than that in the Eskes picture (it is even larger for the $^{1}\!B_1$ case). This results in a stabilization of the $^{1}\!A_1$ state to even more negative $A-\Delta$ or extending even further into the Mott Hubbard regime of the ZSA clasification scheme. 

Figure~\ref{phase_diagram}(b) illustrates the impact on the phase
boundaries of the number of O-2p orbitals kept in the model:
full/dashed lines are for the N7/N9 model. The conventional relations
$t_{pd} \approx \sqrt{3} t_{pd\sigma}/2 = 2t_{pd\pi}$ and $t_{pp\sigma}=0.9$ eV, $t_{pp\pi}=0.2$ eV are used for
the N9 model. Clearly, adding the second in-plane O-2p orbital in the
model does not have significant effects on the phase boundaries,
except to sligthly shift the I-III boundary. The same is true if the
$p_z$ orbitals are also included, in N11 (not shown). For comparison,
the black line denotes the critical $A-\Delta$ for the appearance of
low-energy bound state of Zhang-Rice singlet nature in the N3 model.
At larger $t_{pd}$ this agrees well with the $^{1}\!A_1$ boundary for
N7 model, suggesting minor differences there between the N3 and N7
models. 

Two-dimensional phase diagrams like those of
Figure~\ref{phase_diagram} may be expected to change depending on
whether the $A-\Delta$ axis is spanned by changing $A$ while keeping $\Delta$
constant, or by changing $\Delta$ while keeping $A$ constant, or by some
other protocol. In Fig.~\ref{phase_diagram3d}
we show how the phase diagram evolves with the charge transfer energy
$\Delta$. The rather weak dependence of the $A-\Delta$ {\em vs.} $t_{pd}$ phase
boundaries upon $\Delta$ confirms the importance of the energy separation
between $A$ and $\Delta$. Specifically, as $\Delta$ governs the energy difference between the $d^9$ and $d^{10}L$ state, $A-\Delta$ governs the average energy difference between $d^8$ and $d^9L$. If $A$ is less than $\Delta$, we are closer to a Mott-Hubbard limit than a charge-transfer gap limit. In that case, the $d^8$ triplet is the lowest energy electron removal state as clearly seen in Fig.~\ref{phase_diagram} although the singlet lowest energy state extends well into this negative $A-\Delta$ region.

\begin{figure}[b]
\psfig{figure=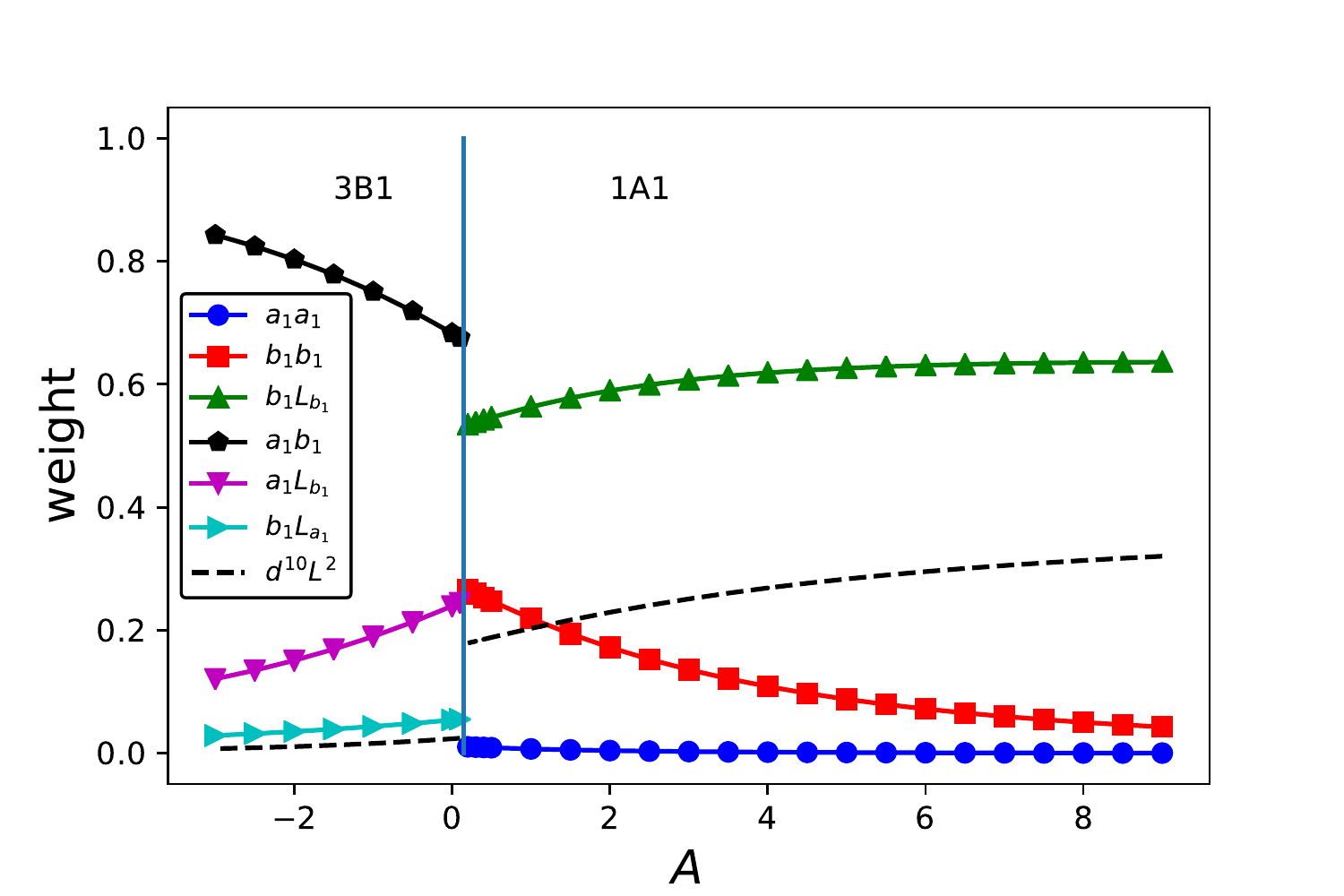}.pdf,height=5.5cm,width=8.0cm,angle=0,clip}
\caption{(Color online) Variation of the ground state weights of the
  dominant components versus $A$, for fixed  $t_{pd}=1.3$ eV,
  $\Delta=3.5$ eV. The vertical line denotes the critical value $A=0.1$ eV separating
  the two phases.}
\label{GSweights}
\end{figure}

To further characterize the evolution of the ground state from region
II ($^{3}\!B_1$) to III ($^{1}\!A_1$), Fig.~\ref{GSweights} plots how
the weights of the dominant components to the ground state change with
$A$, for realistic values of $t_{pd}=1.3$ eV, $\Delta=3.5$ eV. As expected (see
also Eq. ~\ref{eq:HM}), in region III the ground state is dominated by
the $b_1 L_{b_1}$ singlet, which is the equivalent of the ZRS. In
region II, the high spin ground state is dominated by the $a_1 b_1$
triplet. However, in both cases there are significant contributions
from other configurations with the correct symmetry. This shows that
overly simplistic models, which project out everything but the largest probability
component, may be qualitatively correct but will certainly not be
quantitatively accurate for realistic values of the parameters.

\begin{figure}[t]
\psfig{figure=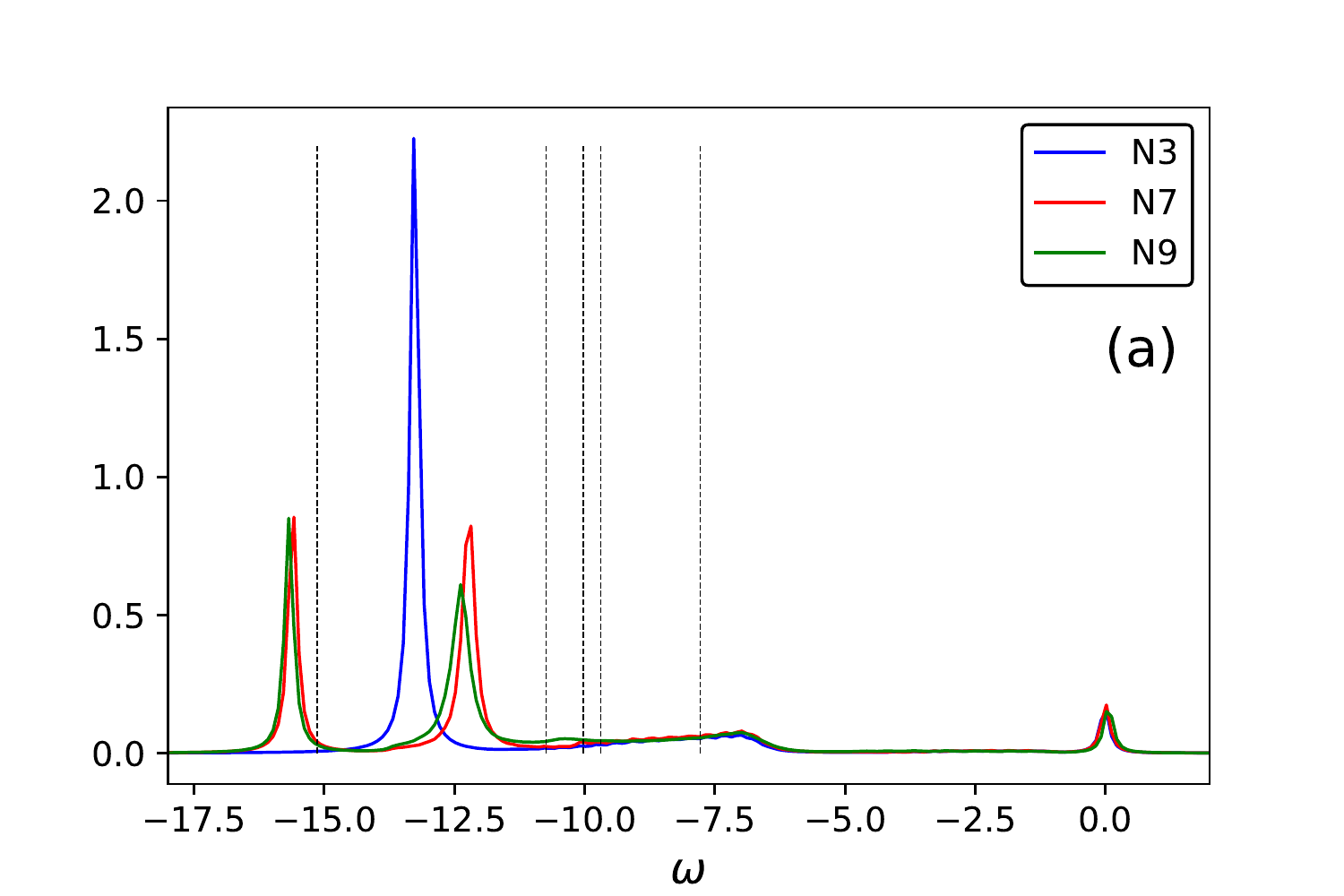}.pdf,height=5.0cm,width=8.0cm,angle=0,clip}
\psfig{figure=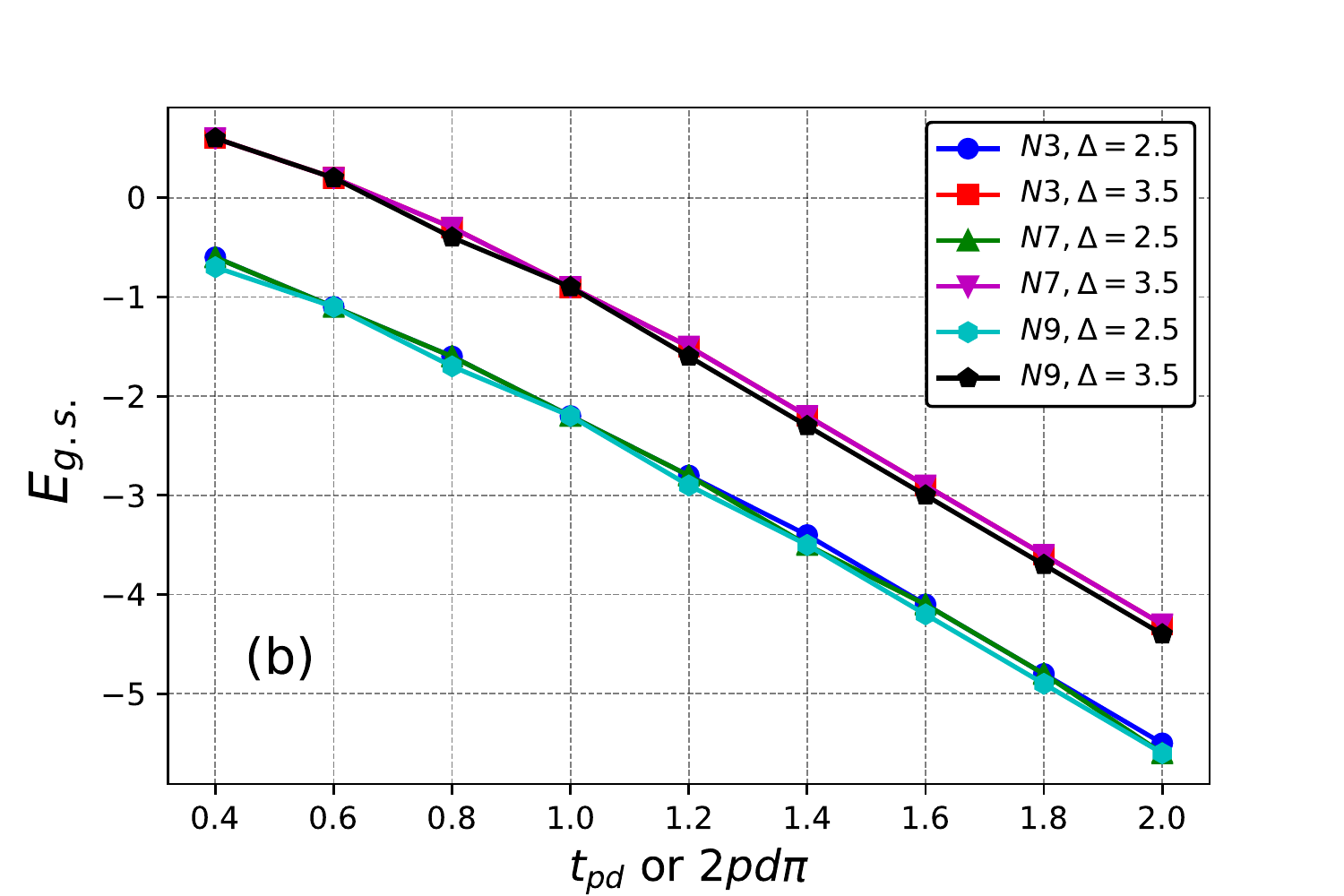}.pdf,height=5.0cm,width=8.0cm,angle=0,clip}
\psfig{figure=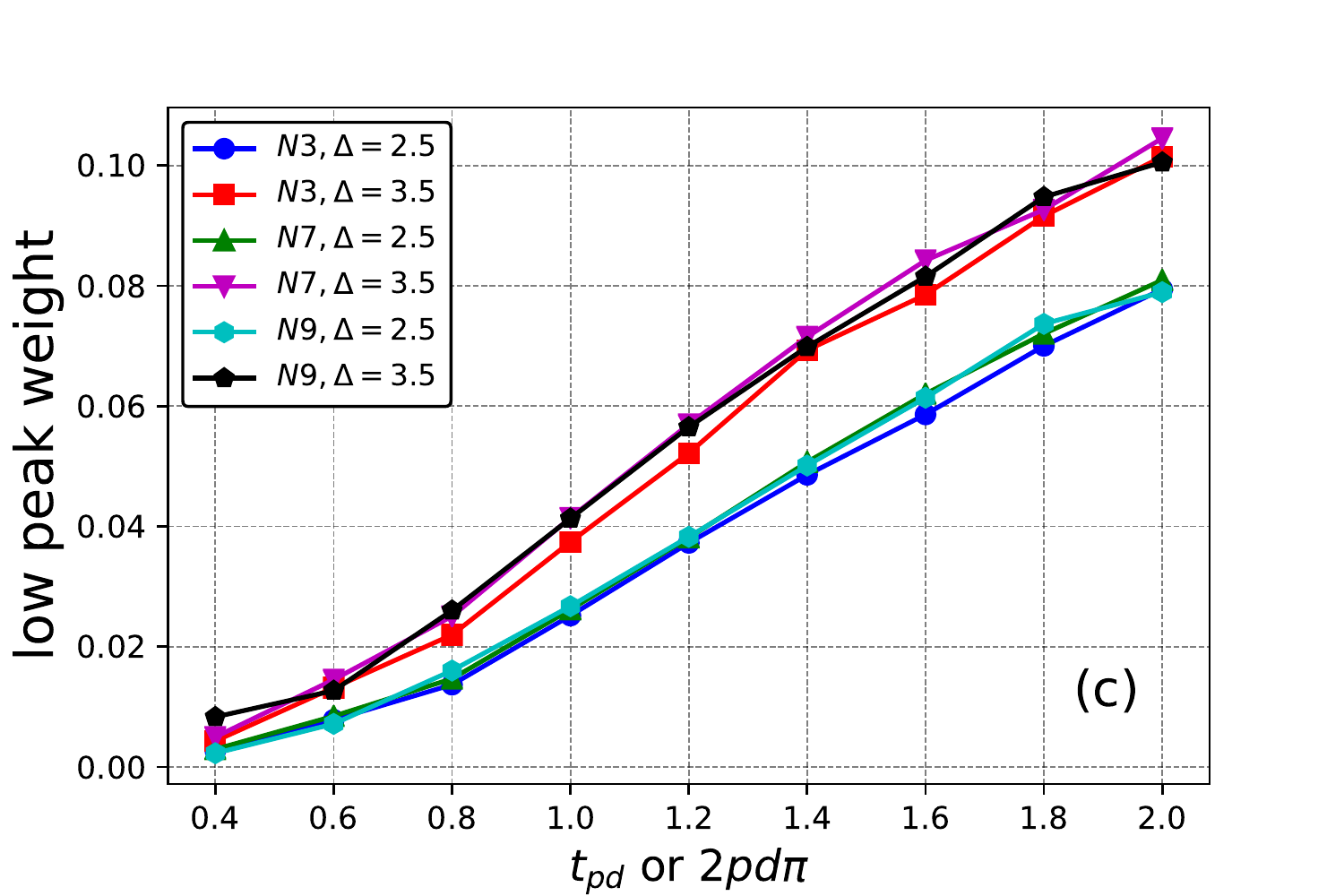}.pdf,height=5.0cm,width=8.0cm,angle=0,clip}
\caption{(Color online) (a) Comparison of two-hole spectra for the N3,
  N7 and N9 models. Parameters are $\Delta=3.0$ eV, $A=6.5$ eV, $U_{dd}=A+4B+3C=8.84$ and $t_{pd}=1.3$ eV, $t_{pp}=0.65$ eV, $t_{pd\sigma}=2.6/\sqrt{3}$ eV, $pd\pi=0.65$ eV, $t_{pp\sigma}=1.0$ eV, $t_{pp\pi}=0.3$ eV; (b-c) the ground-state energy and the
  weight of its corresponding peak as functions of the Cu-O
  hybridization $t_{pd}$ and charge-transfer energy $\Delta$.}
\label{lowpeak_imp}
\end{figure}

Finally, we compare the results of the conventional three-orbital
Emery model (N3 in our notation) against the N7 and N9 results, to see
if the multiplet physics plays any essential role at values of the
parameters believed to be reasonable for cuprates. To achieve this, we
performed the N3 calculation with the same Cu-O hybridization, O-$2p$
hopping integrals, and charge transfer energy $\Delta$ in region III of the
phase diagram as in the N7 model, but keeping only the
$b_1(d_{x^2-y^{2}})$ orbital with a Hubbard-like $U_{dd}=A+4B+3C$ (see
Table~\ref{table1}).

Figure~\ref{lowpeak_imp}(a) compares the spectral weight with
$^{1}\!A_1$ symmetry for the three models. Clearly, the ground-state
peak and the intermediate energy continua due to Cu-O hybridization
are in good agreement. However, the high energy regions ($\omega \approx 10.0$ eV)
of the N7 and N9 models differ from that of the N3 model, as the
latter has a double-occupancy peak at about $U_{dd}$ instead of the
full multiplet spectrum of the former.

Figures~\ref{lowpeak_imp}(b) and (c) focus on the ground-state energy
and peak weight, respectively. We see that the models are in very good
agreement, suggesting that the multiplet physics and/or inclusion of
non-ligand O-2p orbitals has little relevance for the nature of the
ground-state. These results appear to confirm the
validity of the conventional three-orbital Emery model for describing
the low energy physics of the cuprates.

As a final note,  we remark again about the potential imporance of including more than one Cu atoms in addition to explictly including the O states, as done in Lau's previous calculations~\cite{Lau}. This  might lead to very different conclusions especially when considering the effects of doping to a level where the ZR like states strongly overlap, which already occurs at less than 10\% doping.  

\section{Conclusion}\label{Conclusion}

In summary, we used variational exact diagonalization to revisit the
problem of the spectra of two holes doped into an otherwise full
CuO$_{2}$ layer, modelled as a Cu-d$^{10}$ impurity properly embedded into
a square lattice of O-2p$^{6}$. While the relevance of the full Cu
multiplet structure was considered before, with results in qualitative
agreement with ours, the novelty here is that we use a realistic tight
binding band structure for the O band and consider the implications of
adding non-ligand 2p orbitals, as well.

We find that using a realistic O-2p band structure does not change
qualitatively the two-hole spectra in the various symmetry channels,
when compared against those obtained using a featureless, semielliptic
band structure. However, there are significant quantitative changes.
For example, the region in the phase diagram favoring a bound
ground-state with $^{1}\!A_1$ symmetry is enlarged significantly and extends well into the Mott Hubbard region of the ZSA classification sheme. This
proves that using a realistic band-structure has non-trivial
quantitative consequences, that are important if detailed modelling
and comparison to experiments is desired. This is an important
lesson for any impurity-type calculations, including for the use of the dynamical
mean-field theory (DMFT) approximations.
In particular, it is important to include the O-2p states explicitly in the impurity Hamiltonian together with the full multiplet structure, rather than restricting to coupling to a bath of Cu or effective Cu 3d states.

Speaking of approximations, our results also caution against the approach of using Wannier functions together with $A, B, C$ Racah parameters renormalized so as to obtain an atomic limit multiplet similar to the one produced by the strong hybridization. As discussed when analyzing the results from Fig.~\ref{atomic}(a), the two have very different splittings and even ordering of the peaks in the various symmetry channels. For example, in order to get the singlet-triplet crossing within an atomic multiplet approach, one would need to have a Hund's rule $J_d<0$, which is not reasonable. In the original papers describing the effective “screening” of $U$, the lack of screening of $J_h$ and other Hund's rule interactions were obtained with the assumption that the covalency and transition metal to oxygen hybridization would be explicitly included in the model Hamiltonians. This is very different from trying to account for the effects of hybridization through the use of Wannier functions.


Furthermore, we find that the three-orbital Emery model reproduces
well the low-energy results obtained in the $^{1}\!A_1$ symmetry
channel of the N7 and N9 models. In particular, its ground-state is
consistent with the ZRS, but the overlap with the ZRS wavefunction is
only around $50\%$ for reasonable values of the parameters. This
raises questions about the accuracy of projecting the Emery model onto
ZRS, to obtain simple one-band Hamiltonians\cite{Hadi2}. 
We point out again the importance of including all the multiplets when discussing energy scales larger than about 1eV as in many of the optical and photoemission spectroscopies. An obvious example is the appearance of the so called ``waterfall'' feature~\cite{Lanzara} at energies of about 1eV above the lowest energy electron removal state. This can trivially be explained by taking into account all the multiplets and their hybridization with the oxygen bands, forming a broad region in energy where a huge number of bands cross and overlap so that a broad continuum sets in a  momentum distribution plot of ARPES spectroscopy. 

We also find that adding more O-2p orbitals
(in the N9 and N11 models) has essentially no consequences on the $^{1}\!A_1$ symmetry
low-energy spectra. All these results seem to confirm the validity
of the conventional three-orbital Emery model for describing the low
energy physics. However, more care is needed before drawing that
conclusion, as the Emery model completely misses the low-energy peaks
of other symmetries that are revealed by the full calculation, and
which may be relevant to various properties of the cuprates. In fact,
it is worth emphasizing that the projection onto different irreducible
representations is only possible because we treat a single Cu
impurity, as opposed to a lattice of Cu sites. For a lattice, these
various symmetries will mix everywhere in the Brillouin zone except at
high-symmetry points, and thus it is questionable whether these states
with other symmetries are truly irrelevant.
In fact, the study by Lau~\cite{Lau} clearly demonstrates a strong ferromagnetic ordering of the two Cu spins sandwiching an oxygen hole. This is a strong indication that more extended cluster models need to be studied to check whether the influence of the magnetic order and of the hole or electron doping  on the stability of the ZRS in single-band Hubbard model scenario, is indeed valid in the doping region where superconductivity arises. 

The lower symmetry of the lattice (as opposed to an impurity) may also explain how the $z$-axis
polarization, discussed in the introduction, may be accounted for. The
$d_{3z^2-r^2}, d_{xz}$ and $d_{yz}$ orbitals have very little
contribution to the $^{1}\!A_1$ ground-state, but they contribute
significanly to the low-energy peaks in the other symmetry channels. A
lattice calculation that breaks the $D_{4h}$ point group symmetry may
boost not only their contribution to the ground-state, but also the
importance of the O-2p$_z$ orbitals that mostly hybridize with them.

To settle these questions, calculations for the lattice equivalent
of the N7 model are needed. Needless to say, an exact solution is a very hard
challenge. Instead, it may be possible to obtain accurate results
using variational approximations similar to those used here, but
extended to a full Cu lattice. We will investigate this  next.

{\em Acknowledgements:} This work was funded by Stewart Blusson
Quantum Matter Institute, Natural Sciences and Engineering Research Council (NSERC) for Canada, and Canada First Research Excellence Fund (CFREF).

\bibliography{CurroBibliography}

\begin{thebibliography}{40}
\bibitem{ligandfield}
We adopt the values in Figure 2 of Ref.~\cite{Eskes90} for the ligand field splittings. Note that the precise values do not modify the implication in the text.
\end{thebibliography}

\end{document}